\documentclass[twoside,journal,10pt]{IEEEtran}

\usepackage{amsmath}
\usepackage{amsfonts}
\usepackage{amssymb}
\usepackage{setspace}
\usepackage{graphicx}
\usepackage{amsthm}
\usepackage{color}
\usepackage{cite}
\usepackage{bm}
\usepackage[outdir=./fig/]{epstopdf}
\usepackage{psfrag}
\usepackage{multirow}
\usepackage[yyyymmdd,hhmmss]{datetime} \usepackage{epsfig}
\usepackage{psfrag}
\usepackage{pst-node}
\usepackage{fancyhdr,bbm}
\usepackage{color}
\usepackage{float}
\usepackage{comment}
\usepackage{mathtools}

\usepackage{tabularx,booktabs}
\newcolumntype{Z}{@{\hspace{-.8mm}} >{\centering\arraybackslash} X @{\hspace{-.8mm}}}

\usepackage[lined,boxed]{algorithm2e}
\SetAlCapSkip{1.5mm}
\SetAlFnt{\footnotesize}
\SetAlCapFnt{\footnotesize}
\SetAlCapNameFnt{\footnotesize}

\usepackage[columnwise,switch,mathlines]{lineno}

\newcommand{\be}{\begin{equation}}
\newcommand{\ee}{\end{equation}}
\newcommand{\ist}{\hspace*{.3mm}}
\newcommand{\rmv}{\hspace*{-.3mm}}

\newcommand{\cl}[1]{\mathcal{#1}}
\newcommand{\nn}{\nonumber}
\newcommand{\T}{\text{T}}

\newcommand{\V}[1]{\bm{#1}}
\newcommand{\Set}[1]{\mathcal{#1}}

\newcommand{\deq}{\triangleq}

\definecolor{FmCorrection}{RGB}{0,120, 0}
\newcommand{\rd}{\textcolor{black}}

\allowdisplaybreaks

\newlength{\figureheight}
\newlength{\figurewidth}
\definecolor{myYellow}{HTML}{edb120}
\definecolor{myRed}{HTML}{d95319}
\definecolor{myBlue}{HTML}{4dbeee}
\definecolor{myGreen}{HTML}{77ac30}
\definecolor{myPurple}{HTML}{7e2f8e}

\providecommand{\rd}{}
 
\definecolor{BLUE}{rgb}{0,0,1}

\newdateformat{monthyeardate}{\monthname[\THEMONTH] \THEDAY, \THEYEAR}

\newcommand{\paperTitle}{Classification-Aided Multitarget Tracking \\ Using the Sum-Product Algorithm\vspace*{2.5mm}}
\newcommand{\paperTitleMarkboth}{}

\floatstyle{boxed}
\newfloat{floatbox}{thp}{lob}[section]
\floatname{floatbox}{Text Box}

\excludecomment{comment}

\mathtoolsset{showonlyrefs = true}

\begin{document}

\title{\paperTitle}

\author{

\vspace{-2mm}
		Domenico Gaglione*, 		Giovanni Soldi*, 
		Paolo Braca,~\IEEEmembership{Senior Member,~IEEE},
		Giovanni De Magistris, \\
 		Florian Meyer,~\IEEEmembership{Member,~IEEE},
		and Franz Hlawatsch,~\IEEEmembership{Fellow,~IEEE\vspace{-.1mm} }
		
\vspace{-1.5mm}    

\thanks{
This work was supported in part by the NATO Allied Command Transformation (ACT) under the DKOE project
and by the Austrian Science Fund (FWF) under grants J 3886-N31 and 
P 32055-N31.
D.\ Gaglione, G.\ Soldi, P.\ Braca, and G.\ De\ Magistris are with the NATO Centre for Maritime Research and Experimentation (CMRE), La~Spezia, Italy (e-mail: [domenico.gaglione, giovanni.soldi, paolo.braca, giovanni.demagistris]@cmre.nato.int). 
F.\ Meyer 
is with the Scripps Institution of Oceanography and the Electrical and Computer Engineering Department, University of California San Diego, La Jolla, CA, USA (e-mail: flmeyer@ucsd.edu).
F.\ Hlawatsch is with the Institute of Telecommunications, TU Wien, Vienna, Austria (e-mail: franz.hlawatsch@tuwien.ac.at).}
\thanks{* Co-first authors.}
}

\newpage\null\thispagestyle{empty}
\setcounter{page}{0}

{\huge IEEE Copyright Notice}

\vspace{4mm}

\begin{minipage}{\textwidth}
\noindent\large
\textcopyright 2020 IEEE. Personal use of this material is permitted. Permission from IEEE must be obtained for all other uses, in any current or future media, including reprinting/republishing this material for advertising or promotional purposes, creating new collective works, for resale or redistribution to servers or lists, or reuse of any copyrighted component of this work in other works.

\vspace{10mm}

\Large
Accepted to be published in IEEE Signal Processing Letters.

\end{minipage}

\newpage

\maketitle

\begin{abstract}
Multitarget tracking (MTT) is a challenging task that aims at estimating the number of targets and their states from measurements 
of the target states provided by one or multiple sensors.
Additional information, 
such as imperfect estimates of 
target classes provided
by a classifier, can 
facilitate the target-measurement association and thus improve 
MTT 
performance.
In this letter, we describe how a recently proposed MTT framework based on the sum-product algorithm 
can be 
extended to efficiently exploit 
class information.
The effectiveness of the proposed approach is demonstrated 
by simulation results.
\vspace{-.1mm}
\end{abstract}

\begin{IEEEkeywords}
Multitarget tracking, 
probabilistic data association, 
sum-product algorithm, 
classification,
factor graph.
\vspace{-.5mm}
\end{IEEEkeywords}

\section{Introduction}
\label{sec:introduction}
Multitarget tracking (MTT) 
\cite{BarWilTia:B11,Mah:B07}
aims at estimating the number of targets and their states
from measurements provided by one or multiple sensors.
A major challenge 
in MTT 
is 
posed by
measurement origin uncertainty (MOU) \cite{BarWilTia:B11}, i.e., the fact that it is not known 
if a measurement is produced by a target, and by which target.
A Bayesian message passing algorithm
that efficiently 
addresses MOU was presented in~\cite{WilLau:J14}. This algorithm has been used to develop an 
MTT tracking method
\cite{MeyBraWilHla:J17,MeyKroWilLauHlaBraWin:J18} that
employs 
the sum-product algorithm (SPA) \cite{KscFreLoe:01,Loe:04,LoeDauHuKorPinKsc:J07} 
to 
approximate the marginal posterior probability density functions (pdfs) of the target states.
SPA-based MTT
was demonstrated to be highly scalable 
and to outperform several previously proposed MTT methods \cite{MeyBraWilHla:J17,MeyKroWilLauHlaBraWin:J18}.
Moreover, 
the flexibility
of the SPA approach
allows for several extensions 
such as integration of data provided by heterogenous sensors \cite{GaglioneBS:C18,SoldiGMHBFW:C19} and adaptation to time-varying 
parameters 
\cite{SoldiMBH:J19}.

Compared to processing only the 
sensor measurements, as is done by most existing MTT methods 
(e.g., \rd{\cite{SoldiMBH:J19,MeyBraWilHla:J17,MeyKroWilLauHlaBraWin:J18,GarWilGranSven:J18,XiaGranSvenGar:C17,GranSvenXiaWilGar:C18, 
VerGodPer:J05,VoSinDou:J05,VoVoCan:J07,BatChiMorPapFarGra:J12,VoVoPhu:J14,Wil:J15,NanBloCoaRab:J16,ShaSauBucVar:J19,LiWanLiaPan:J20}}),
exploiting additional target-related information generally leads to improved MTT performance.
Here, we consider the exploitation of imperfect
\textit{class} information \cite{Drummond:C01,BarShalomKirGok:J05_41}, which assigns
a target 
to one among several categories.
In the maritime domain, for example, such categories could be commercial ships, military ships, and fishing boats.
This class information is generally the output of a classifier and allows, e.g., the use of 
class-dependent motion or measurement models.

Feature-aided and classification-aided tracking techniques \cite{Drummond:C01,BarShalomKirGok:J05_41,
RavindraBD09,SinghTDPW09_39,YingZS11_1,GeorgescuW13,
Mellema14,MoriCC14_50} have lately encountered a growing interest, in particular following 
recent advances in deep learning classification methods \cite{HuangZDZ:J12,ChenLZWG:J14,LeCunBH:J15}.
In this letter,
we show how imperfect
target class information can be integrated into the 
SPA-based MTT method of \cite{MeyBraWilHla:J17,MeyKroWilLauHlaBraWin:J18}. 
More concretely, we propose an SPA-based 
MTT method that takes into account the output of a
classifier
distinguishing between target-related classes as well as between target- and clutter-originated measurements.
Our simulation results show that the proposed classification-aided MTT method outperforms the 
MTT method 
of \cite{MeyBraWilHla:J17}.

The remainder of this letter is organized as follows.
Section~\ref{sec:system_model_statistical_formulation} presents the system model and 
its statistical formulation.
Section \ref{sec:proposed_method} describes the proposed method.
Section \ref{sec:simulation_results} reports 
simulation results.

\section{System Model and Statistical Formulation}
\label{sec:system_model_statistical_formulation}

\vspace{.5mm}

\rd{The system model described in this section recalls the one introduced in \cite{MeyBraWilHla:J17}
and, in addition, establishes statistical characterizations of
the target class and of the imperfect class information provided by the classifier.}

\vspace{-1mm}

\subsection{Target Model}
\label{sec:system_model_target}

We consider $K$ \textit{potential targets} (PTs) indexed by $k \! \in \! \Set{K} \deq \! \{ 1,\ldots,K \}$, whose existence at time $n$ is indicated by 
$r_{n,k} \! \in \! \{ 0,1 \}$, i.e., $r_{n,k} \! = \! 1$ if 
PT $k$ exists and $r_{n,k} \! = \rmv 0$ otherwise.
Note that $K$ is an upper limit of
the 
number of actual targets that can be tracked simultaneously, i.e., the (time-varying) number of actual targets need not be known in advance except that it is not larger than $K$. 
(Scalable SPA-based multisensor
MTT methods
with a time-varying number of PTs $K$
are presented in \cite{MeyKroWilLauHlaBraWin:J18}.)
The state of PT $k$ at time $n$ is represented by the vector $\V{x}_{n,k}$ and consists of the PT's position and possibly further parameters, such as the PT's velocity;
a ``dummy'' state is formally considered also if 
$r_{n,k} \! = \rmv 0$.
Each PT $k$ 
belongs to one of $C$ distinct classes $c \! \in \! \{ 1,\ldots,C \}$ at time $n$;
the class of PT $k$ at time $n$ is specified by the class index variable 
$\ell_{n,k} \in \{ 1, \linebreak \ldots, C \}$, which is unknown just as $\V{x}_{n,k}$ and $r_{n,k}$.
We allow $\ell_{n,k}$ to be time-dependent, although in most applications it is constant.
The time evolution of the state of a PT $k$ that exists at times $n \!-\! 1$ and $n$
(i.e., $r_{n-1,k} \rmv=\rmv r_{n,k} \rmv=\! 1$) is modeled as $\V{x}_{n,k} = \V{\theta}_{\ell_{n,k}} \big( \V{x}_{n-1,k},\V{u}^{(\ell_{n,k})}_{n,k} \big)$,
	where $\V{u}^{(\ell_{n,k})}_{n,k}$ is a process noise
that is independent and identically distributed (iid) across $n$ and $k$. 
The state transition function $\V{\theta}_{\ell_{n,k}}(\cdot,\cdot)$ 
is selected from a set ${\{\V{\theta}_{c}(\cdot,\cdot) \}}_{c=1}^C$ by the class index variable $\ell_{n,k}$.
Furthermore, also the statistics of $\V{u}_{n,k}^{(\ell_{n,k})}$ generally depend on $\ell_{n,k}$.
The function $\V{\theta}_{\ell_{n,k}}(\cdot, \cdot)$ and the statistics 
of $\V{u}^{(\ell_{n,k})}_{n,k}$
define the state transition 
pdf $f (\V{x}_{n,k} | \V{x}_{n-1,k},\ell_{n,k})$.
In addition to the dynamics of the PTs, also other PT characteristics
(e.g., color, size, type) 
may be related to the PT class.

For convenience, we 
define the \emph{augmented state} vector $\V{y}_{n,k} \linebreak \deq \! [\V{x}_{n,k}^{\T},r_{n,k},\ell_{n,k}]^{\T}\rmv$ as well as the stacked
\rd{vector
$\V{y} \rmv \deq \rmv [\V{y}_{0}^{\T},\ldots, \linebreak \V{y}_{n}^{\T}]^{\T}$, where $\V{y}_{n} \rmv \deq \rmv [\V{y}_{n,1}^{\T},\ldots,\V{y}_{n,K}^{\T}]^{\T}$.}
\rd{The time evolution of the augmented state of a PT $k$ is statistically described by}
the transition pdf $f(\V{y}_{n,k} | \V{y}_{n-1,k})$.
\rd{Assuming }
that 
$\V{x}_{n,k}$ and $r_{n,k}$ are conditionally 
independent of $\ell_{n-1,k}$ given $\V{x}_{n-1,k}$, $r_{n-1,k}$, and $\ell_{n,k}$,
\rd{this transition pdf can be obtained as}
\begin{align}
	f(\V{y}_{n,k} | \V{y}_{n-1,k}) &= f(\V{x}_{n,k},r_{n,k},\ell_{n,k} | \V{x}_{n-1,k},r_{n-1,k},\ell_{n-1,k}) \nonumber \\
	&=  f(\V{x}_{n,k}, r_{n,k} | \ell_{n,k}, \V{x}_{n-1,k},r_{n-1,k},\ell_{n-1,k}) \nonumber \\
	&\hspace{5mm} \times p (\ell_{n,k} |\V{x}_{n-1,k},r_{n-1,k},\ell_{n-1,k})\nonumber \\
	&=  f(\V{x}_{n,k}, r_{n,k} | \ell_{n,k}, \V{x}_{n-1,k},r_{n-1,k}) \nonumber \\
	&\hspace{5mm} \times p (\ell_{n,k} |\V{x}_{n-1,k},r_{n-1,k},\ell_{n-1,k}) .
	\label{eq:augmented_state_transition_pdf} 
\end{align}
Here, an expression 
of $f(\V{x}_{n,k}, r_{n,k} | \ell_{n,k}, \V{x}_{n-1,k}, r_{n-1,k})$ 
is provided in \cite[Sec. II-C]{SoldiMBH:J19}. 
If the PT's dynamic model is independent of 
$\ell_{n,k}$,
then $f( \V{x}_{n,k}, r_{n,k} | \ell_{n,k},\V{x}_{n-1,k}, r_{n-1,k} ) \! = \! f( \V{x}_{n,k}, r_{n,k} | \V{x}_{n-1,k}, r_{n-1,k})$;
an expression of this pdf is provided in \cite[Sec. II-A]{MeyBraWilHla:J17}.
Furthermore, the transition probability mass function (pmf)
$p(\ell_{n,k} | \V{x}_{n-1,k}, r_{n-1,k}, \ell_{n-1,k})$ is 
given as follows. If PT $k$ did not exist at time $n \!-\! 1$, i.e., $r_{n-1,k} \! = \!0$, then  $p(\ell_{n,k} | \V{x}_{n-1,k},0, \ell_{n-1,k}) \! = \! 1 / C$. 
Else,
if PT $k$ 
existed at time $n \!-\!1$, i.e., $r_{n-1,k} \!=\! 1$, then $p(\ell_{n,k} | \V{x}_{n-1,k}, 1,\ell_{n-1,k})$
is described by the transition matrix
\rd{$\V{D}(\V{x}_{n-1,k}) \! \in \! [0,1]^{C \times C}$,} with
\rd{$[\V{D}(\V{x}_{n-1,k})]_{i,j} \! = \!  p(\ell_{n,k} \!\rmv = \! i | \V{x}_{n-1,k},1,\ell_{n-1,k} \! = \! j)$, $i,j \! \in \! \{ 1,\ldots,C \}$.}
We 
note that
$\sum_{i = 1}^C \rd{[\V{D}(\V{x}_{n-1,k})]_{i,j}} = 1$.

\vspace{-1mm}

\subsection{Measurement and Classifier Model}
\label{subsec:measurement_model_conf_matrix}

There are $S$ sensors indexed by $s \! \in \! 
\{ 1,\dots,S \}$.
Each sensor $s$ provides,
at time $n$, $M^{(s)}_n$ measurements $\V{q}^{(s)}_{n,m}\ist$, $m \! \in \! \cl{M}^{(s)}_n \linebreak \triangleq \! \{ 1,\dots,M^{(s)}_n \}$.
An existing PT $k$ (i.e., with $r_{n,k} \! = \! 1$) is detected by sensor $s$---in the sense that it generates a measurement at sensor $s$---with probability $P_{\text{d}}^{(s)}(\V{x}_{n,k},\ell_{n,k}) \rmv$.
A measurement
originating from
PT $k$ follows the measurement model $\V{q}_{n,m}^{(s)} \! = \! \V{\psi}_{s}(\V{x}_{n,k},\V{v}_{n,m}^{(s)})$.
Here, $\V{v}_{n,m}^{(s)}$ is 
measurement noise that is iid across $n$ and $m$ and independent across $s$.
The function $\V{\psi}_{s}(\cdot,\cdot)$
and the statistics of $\V{v}_{n,m}^{(s)}$ define the likelihood function $f\big( \V{q}_{n,m}^{(s)} \big| \V{x}_{n,k} \big)$.
A ``false alarm'' (clutter) measurement 
is distributed according to 
pdf $f_{0} \big( \V{q}_{n,m}^{(s)} \big)$.
The number of false alarms at sensor $s$ is Poisson distributed with mean $\mu^{(s)}\rmv$.

Each measurement $\V{q}_{n,m}^{(s)}$ is accompanied by an estimate $\zeta_{n,m}^{(s)}$ of the class index,
which is provided by a classifier.
Here, $\zeta_{n,m}^{(s)} \! = \! 0$ 
expresses the classifier's
belief that measurement $\V{q}_{n,m}^{(s)}$
is clutter-generated, and 
$\zeta_{n,m}^{(s)} \! = \! c \! \in \! \{ 1,\ldots,C \}$ 
that 
it is 
generated 
by a target that belongs to class $c$.
For convenience, we define the \textit{augmented measurement} vector 
$\V{z}_{n,m}^{(s)} \! \triangleq$\linebreak $[\V{q}_{n,m}^{(s)\T},\zeta_{n,m}^{(s)}]^{\T}$
as well as the stacked vectors
$\V{z}^{(s)}_n \! \triangleq \! \big[ \V{z}^{(s)\T}_{n,1},\ldots,$\linebreak $\V{z}^{(s)\T}_{n,M^{(s)}_n} \big]^{\T}$, 
$\V{z}_n \! \triangleq \! \big[ \V{z}^{(1)\T}_n \rmv, \ldots, \V{z}^{(S)\T}_n \big]^{\T}$, 
and $\V{z} \! \triangleq \! [\V{z}^\T_1,\ldots,\V{z}^\T_n]^{\T}$.

The statistical dependency of 
a target-generated augmented measurement 
$\V{z}_{n,m}^{(s)}$ on the underlying PT state $\V{x}_{n,k}$ and PT class $\ell_{n,k}$ is described by the likelihood function $f \big( \V{z}_{n,m}^{(s)} \big| \V{x}_{n,k}, \ell_{n,k} \big)$.
Assuming that
$\zeta_{n,m}^{(s)}$ 
is conditionally independent of $\V{q}_{n,m}^{(s)}$ given
$\V{x}_{n,k}$ and 
$\ell_{n,k}$, 
we obtain
\begin{align}
	f\big( \V{z}_{n,m}^{(s)} & \big| \V{x}_{n,k}, \ell_{n,k} \big) = f\big( \V{q}_{n,m}^{(s)}, \zeta_{n,m}^{(s)} \big| \V{x}_{n,k}, \ell_{n,k} \big) \nonumber \\ 
	&\hspace{1mm}= p \big(\zeta_{n,m}^{(s)} \big|\V{q}_{n,m}^{(s)}, \V{x}_{n,k}, \ell_{n,k} \big)
			f \big( \V{q}_{n,m}^{(s)} \big| \V{x}_{n,k}, \ell_{n,k} \big) \nonumber \\[0mm]
	& \hspace{1mm}= p \big( \zeta_{n,m}^{(s)} \big| \V{x}_{n,k}, \ell_{n,k} \big) \ist f \big( \V{q}_{n,m}^{(s)} \big| \V{x}_{n,k}, \ell_{n,k} \big)  \nonumber \\[0mm]
	& \hspace{1mm}= p \big( \zeta_{n,m}^{(s)} \big| \V{x}_{n,k}, \ell_{n,k} \big)  \ist f \big( \V{q}_{n,m}^{(s)} \big| \V{x}_{n,k} \big).
		\label{eq:factorisation_likelihood_ais_single_cluster} \\[-6mm]
	\nonumber
\end{align}
Here, in the
last step, we used 
that the measurement model $\V{\psi}_s (\cdot\ist,\cdot)$ does not depend on 
$\ell_{n,k}$. 
The pmf $p \big( \zeta_{n,m}^{(s)} \big| \V{x}_{n,k},\ell_{n,k} \big)$ models the 
performance of the classifier;
it is described by a
confusion matrix
\rd{$\V{G}^{(s)}(\V{x}_{n,k}) \! \in \! [ 0,1 ]^{(C + 1) \times C}$.}
Here,
\rd{ $[\V{G}^{(s)}(\V{x}_{n,k})]_{i,j} \rmv=\rmv p \big( \zeta_{n,m}^{(s)} \!=\! i \big| \V{x}_{n,k}, \ell_{n,k} \! = \! j \big)$}
is the 
probability that
the classifier output is $i \! \in \! \{ 0,\ldots,C \}$ when the measurement is generated by a target belonging to class
$j \! \in\! \{ 1, \ldots, C \}$.
We note that
 $\sum_{i=0}^C \rd{[\V{G}^{(s)}(\V{x}_{n,k})]_{i,j}} \rmv=\rmv 1$.

The pdf of
a false alarm (clutter-generated) augmented measurement is
given by 
$f_{\text{FA}} \big( \V{z}_{n,m}^{(s)} \big) \rmv= \rmv   p_{0}\big( \zeta_{n,m}^{(s)} \big| \V{q}_{n,m}^{(s)} \big) \ist f_{0} \big( \V{q}_{n,m}^{(s)} \big)$, 
where $p_{0}\big( \zeta_{n,m}^{(s)}\!=\!i \big|\V{q}_{n,m}^{(s)}\big)$ is the probability conditioned on $\V{q}_{n,m}^{(s)}$ that the classifier output is 
$i \! \in \! \{0,\ldots,C \}$ when the measurement $\V{q}_{n,m}^{(s)}$ is clutter-generated.

We 
note that the proposed model can 
be 
extended to 
classifiers providing ``soft'' probabilistic information.
Here, instead of a class estimate $\zeta_{n,m}^{(s)}$, the classifier output is
a probability vector $\V{p}_{n,m}^{(s)} \! = \! \big[ p_{n,m,1}^{(s)}, \ldots, p_{n,m,C}^{(s)} \big]^{\T}\rmv$ 
where 
$p_{n,m,i}^{(s)} \! \in \! [0,1]$ 
is the classifier's estimate of the probability 
that measurement $m$
is generated by a target belonging to class $i \in \{1,\ldots,C\}$.

\vspace{-1.5mm}

\subsection{MOU Model}
\label{subsec:MOU}
The association between the measurements  and the existing PTs is 
unknown, and it is also possible that a measurement $\V{q}_{n,m}^{(s)}$ 
does not originate from any existing PT (false alarm) or 
an existing PT 
does not generate any measurement (missed detection).
We make the assumption---known 
as point target assumption---that an existing PT can generate at most one measurement at a given sensor and a measurement can originate from at most one existing PT \cite{BarWilTia:B11}.
Let us define the \emph{PT-oriented association variable} $a^{(s)}_{n,k}$, $k \! \in \! \mathcal{K}$ to be $m \! \in \! \mathcal{M}_{n}^{(s)}$ if PT $k$ generates measurement $m$ at sensor $s$, and zero if PT $k$ is missed by sensor $s$.
Similarly, 
the \emph{measurement-oriented association variable} $b^{(s)}_{n,m}$, $m \! \in \! \mathcal{M}_{n}^{(s)}$ is $k \! \in \! \mathcal{K}$ if measurement $m$ at sensor $s$ originates from PT $k$, and zero if it is a false alarm.
Following \cite{BayShaSha:J08_54,WilLau:J14}, we define the indicator function $\Psi_{km}^{(s)}\big(a_{n,k}^{(s)}, b_{n,m}^{(s)}\big)$ to be one if the values of $a_{n,k}^{(s)}$ and $b_{n,m}^{(s)}$ are consistent, i.e., if they do not
describe different PT-measurement associations, and zero otherwise.
More formally, $\Psi_{km}^{(s)}\big(a_{n,k}^{(s)}, b_{n,m}^{(s)} \big) \! = \! 0$ if 
either $a_{n,k}^{(s)} \! = \! m$ and $b_{n,m}^{(s)} \! \neq \! k$ or $a_{n,k}^{(s)} \! \neq \! m$ and $b_{n,m}^{(s)} \! =\! k$, and $\Psi_{km}^{(s)}\big(a_{n,k}^{(s)},b_{n,m}^{(s)} \big) \! = \! 1$ otherwise.
The vectors $\V{a}$ and $\V{b}$ 
comprise, respectively, all the $a^{(s)}_{n'\rmv,k}$, $k \!\in\! \mathcal{K}$ and all the $b^{(s)}_{n'\rmv,m}$, $m \!\in\! \mathcal{M}_n^{(s)}$ 
for all the sensors $s$ and all
the times $n' \! = \! 1, \ldots, n$.

A Bayesian network showing the statistical dependencies among $\V{y}$, $\V{a}$, $\V{b}$,
$\V{z}$, and $M^{(s)}_{n'}$ for all sensors $s \! \in \! \{ 1, \ldots, S \}$ and all times $n' \! = \! 1, \ldots, n\ist$
is presented in the supplementary material manuscript \cite{SupplMat}.

\vspace{-1mm}

\section{The Proposed Method}
\label{sec:proposed_method}

MTT aims to determine if a PT $k \! \in \! \mathcal{K}$ exists and, if it exists,\linebreak to
estimate 
its state $\V{x}_{n,k}$.
This essentially amounts to calculating the posterior existence probability $p(r_{n,k} \! = \! 1 | \V{z})$ and the posterior state pdf $f(\V{x}_{n,k} | r_{n,k} \! = \! 1, \V{z})$.
PT $k$ is detected---i.e., declared to exist---if $p(r_{n,k} \! = \! 1| \V{z})$ is larger than a suitably chosen threshold $P_{\text{th}}$ \cite[Ch.~2]{Poo:B94}. 
Then, 
an estimate of $\V{x}_{n,k}$ is given
by
$\hat{\V{x}}_{n,k} \! \triangleq \! \int \rmv \V{x}_{n,k} \ist f(\V{x}_{n,k} | r_{n,k} \! = \! 1, \V{z} ) \ist \mathrm{d}\V{x}_{n,k}$ \cite[Ch.~4]{Poo:B94}.
The statistics $p(r_{n,k} \! = \! 1| \V{z})$ and $f(\V{x}_{n,k} | r_{n,k} \! = \! 1, \V{z} )$ can be obtained from the posterior pdf $f(\V{x}_{n,k}, r_{n,k}, \ell_{n,k} | \V{z}) \! = \! f(\V{y}_{n,k} | \V{z})$ 
essentially via marginalization. Thus, it remains
to calculate the posterior pdfs $f(\V{y}_{n,k} | \V{z})$ for all $k \!\in\! \mathcal{K}$.

The posterior pdf 
$f(\V{y}_{n,k} | \V{z})$ is a marginal density of the joint posterior pdf $f(\V{y},\V{a},\V{b} | \V{z})$.
With the assumptions made in Section \ref{sec:system_model_statistical_formulation}
and those made in \cite{MeyBraWilHla:J17} (also stated in the supplementary material manuscript \cite{SupplMat})\rd{,
one} can show 
that
\begin{align}
	\nonumber \\[-5.5mm]
	&f(\V{y},\V{a},\V{b}|\V{z}) \propto \prod_{k=1}^K\rmv  f(\V{y}_{0,k}) \rmv \prod_{n'=1}^n \! f(\V{y}_{n'\!,k}|\V{y}_{n'-1,k} ) \nonumber \\[-1.5mm]
	&\hspace{1mm} \times \prod_{s=1}^S  \ist \upsilon^{(s)} \big(\V{x}_{n'\!,k},r_{n'\!,k}, \ell_{n'\!,k}, a_{n'\!,k}^{(s)}; \V{z}_{n'}^{(s)}\big)
			\prod_{m=1}^{M_{n'}^{(s)}} \! \Psi\big(a_{n'\!,k}^{(s)},b_{n'\!,m}^{(s)}\big) \ist. \nonumber \\[-3mm]
	\label{eq:factorization} \\[-6mm]
	\nonumber
\end{align}
Here, $f(\V{y}_{n,k}|\V{y}_{n-1,k})$ is given by \eqref{eq:augmented_state_transition_pdf},
\rd{and $f (\V{y}_{0,k})$ is the prior pdf of the augmented state of PT $k$ at time $n \! = \! 0$, which generally includes also some prior probabilistic information on the PT class.}
Furthermore, 
$\upsilon^{(s)} \big(\V{x}_{n,k}, r_{n,k}, \ell_{n,k}, a_{n,k}^{(s)}; \V{z}_n^{(s)}\big)$ is given for $r_{n,k} \! = \! 1$ by
$P_{\text{d}}^{(s)}(\V{x}_{n,k},\ell_{n,k}) \ist f\big( \V{z}^{(s)}_{n,m} \big| \V{x}_{n,k}, \ell_{n,k} \big)\linebreak  / \big( \mu^{(s)} f_{\text{FA}}\big( \V{z}^{(s)}_{n,m} \big) \big)$
if $a_{n,k}^{(s)} \! = \! m \! \in \! \cl{M}_{n}^{(s)}$ and 
$1 \!-\! P_{\text{d}}^{(s)}(\V{x}_{n,k},\ell_{n,k})$ if $a_{n,k}^{(s)} \! = \! 0$, 
\vspace{-1mm}
and for $r_{n,k} \! = \! 0$ by 
$\delta_{a_{n,k}^{(s)},0}\ist$,
where $f\big( \V{z}^{(s)}_{n,m} \big| \V{x}_{n,k}, \linebreak \ell_{n,k} \big)$ is given by \eqref{eq:factorisation_likelihood_ais_single_cluster} and 
\vspace{-.8mm}
$\delta_{a^{(s)}_{n,k},0}$ is 
one if $a^{(s)}_{n,k} \! = \! 0$ and zero otherwise.
\rd{A detailed derivation of \eqref{eq:factorization} is provided in \cite{MeyBraWilHla:J17};
the resulting factorization
formally differs from \eqref{eq:factorization} only by
the definitions of the transition pdf $f(\V{y}_{n,k} | \V{y}_{n-1,k})$ and the function
$\upsilon^{(s)}\big(\V{x}_{n,k},r_{n,k},\ell_{n,k},a_{n,k}^{(s)}; \V{z}_{n}^{(s)}\big)$.
A sketch of that derivation is included in the supplementary material manuscript \cite{SupplMat}.
If some of the sensors are not accompanied by a classifier providing estimates of the target class,
then for these sensors
the 
definition of the function $\upsilon^{(s)}(\cdot)$ from \cite{MeyBraWilHla:J17} has to be used.}

\begin{figure}[t!]
\centering

\psfrag{f1}[c][c][.81]{\raisebox{0mm}{\hspace{0mm}$f_1$}}
\psfrag{fk}[c][c][.81]{\raisebox{0mm}{\hspace{0mm}$f_K$}}

\psfrag{ft1}[c][c][.81]{\raisebox{0mm}{\hspace{0mm}$\tilde{f}_1$}}
\psfrag{ftk}[c][c][.85]{\raisebox{0mm}{\hspace{0mm}$\tilde{f}_K$}}

\psfrag{y1}[c][c][.81]{\raisebox{-2mm}{\hspace{.4mm}$\V{y}_1$}}
\psfrag{yk}[c][c][.81]{\raisebox{-2mm}{\hspace{.7mm}$\V{y}_K$}}

\psfrag{v1}[c][c][.81]{\raisebox{0mm}{\hspace{0mm}$\upsilon_1$}}
\psfrag{vk}[c][c][.81]{\raisebox{0mm}{\hspace{0mm}$\upsilon_K$}}

\psfrag{a1}[c][c][.81]{\raisebox{0mm}{\hspace{0mm}$a_1$}}
\psfrag{ak}[c][c][.81]{\raisebox{0mm}{\hspace{0mm}$a_K$}}

\psfrag{b1}[c][c][.81]{\raisebox{0mm}{\hspace{0mm}$b_1$}}
\psfrag{bm}[c][c][.81]{\raisebox{0mm}{\hspace{0mm}$b_M$}}

\psfrag{p11}[c][c][.81]{\raisebox{3mm}{\hspace{0mm}$\Psi_{11}$}}
\psfrag{p1m}[c][c][.81]{\raisebox{0mm}{\hspace{-6mm}$\Psi_{1M}$}}
\psfrag{pk1}[c][c][.81]{\raisebox{2.5mm}{\hspace{-6mm}$\Psi_{K1}$}}
\psfrag{pkm}[c][c][.81]{\raisebox{-3mm}{\hspace{0mm}$\Psi_{KM}$}}

\psfrag{s1}[c][c][.81]{\raisebox{-4mm}{\hspace{-7mm}$s \! = \! 1$}}
\psfrag{s2}[c][c][.81]{\raisebox{-4mm}{\hspace{-7mm}$s \! = \! S$}}

\includegraphics[scale=.68]{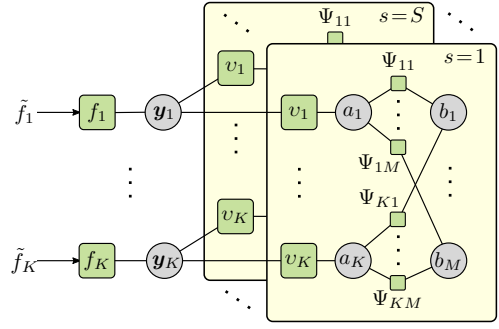}

\vspace{-.5mm}

\caption{Factor graph describing the factorization of $f(\V{y}, \V{a},\V{b} | \V{z})$ in \eqref{eq:factorization} for one time step $n$. For simplicity, the sensor index $s$ and the time index $n$ are omitted, and the following short notations are used:
$\tilde{f}_{k} \! \deq \tilde{f}(\V{y}_{n,k})$,
$f_{k} \! \triangleq \! f(\V{y}_{n,k}|\V{y}_{n - 1,k} )$,
$\upsilon_{k} \! \triangleq \! \upsilon^{(s)} \big(\V{y}_{n,k}, a_{n,k}^{(s)}; \V{z}_n^{(s)}\big)$,
$a_{k} \! \deq \! a_{n,k}^{(s)}$,
$b_{m} \! \deq \! b_{n,m}^{(s)}$,
$\Psi_{km} \! \triangleq \! \Psi_{km}\big(a_{n,k}^{(s)},b_{m,n}^{(s)}\big)$,
and $M \! \triangleq \! \rmv M_n^{(s)}\rmv$.}
\label{fig:factorGraphMS}

\end{figure}

The factor graph \cite{KscFreLoe:01,Loe:04, LoeDauHuKorPinKsc:J07} describing the factorization \eqref{eq:factorization} is shown for one time step in Fig.~\ref{fig:factorGraphMS}.
Following the approach in \cite{MeyBraWilHla:J17}, approximations of the marginal posterior pdfs $f(\V{y}_{n,k} | \V{z})\! = \! f(\V{x}_{n,k}, r_{n,k}, \ell_{n,k} |\V{z})$, known as \emph{beliefs}
and denoted as $\tilde{f}(\V{y}_{n,k}) \! = \! \tilde{f}(\V{x}_{n,k},r_{n,k},\ell_{n,k})$, 
can be calculated 
efficiently
by running iterative SPA message passing on
this factor graph.
Since the 
factor graph contains loops, there is no unique order of calculating the 
messages, and 
different orders may result in different 
beliefs. 
We define the order 
by the following 
rules: first, 
messages are not sent backward in time, and second, \emph{iterative} message passing is only performed for probabilistic data association, and 
separately at each time step and at each sensor.
The second rule implies that for loops involving different sensors, only a single message passing iteration is performed.
The structure of the factor graph in Fig.\ \ref{fig:factorGraphMS} equals that in \cite{MeyBraWilHla:J17}, even though 
the underlying Bayesian model---i.e., the functions 
represented by the factor nodes ``$f_{k}$'' and ``$\upsilon_{k}$'' and the 
variables represented by the variable nodes ``$\V{y}_{k}$''---are different.
The derivation and expressions of the messages are thus analogous to 
\cite{MeyBraWilHla:J17} and
are omitted
because of space restrictions.
However, a detailed statement 
of our method
is provided in the supplementary material manuscript \cite{SupplMat}.

\vspace{-1.5mm}

\section{Numerical Study}\label{sec:simulation_results}

\subsection{Simulation Setup}\label{subsec:basic_sim_setup}

\begin{figure}[!t]

\centering

\psfrag{xla}[c][c][.70]{\raisebox{-6.5mm}{\hspace{0mm}$x_1 \, \text{[m]}$}}
\psfrag{yla}[c][c][.70]{\raisebox{0mm}{\hspace{0mm}$x_2 \, \text{[m]}$}}

\psfrag{A1}[c][c][.70]{\raisebox{0mm}{\hspace{4mm}A}}
\psfrag{A2}[c][c][.70]{\raisebox{-3mm}{\hspace{2mm}\textcolor{myRed}{B}}}
\psfrag{A3}[c][c][.70]{\raisebox{-3mm}{\hspace{-3mm}\textcolor{myYellow}{C}}}
\psfrag{A4}[c][c][.70]{\raisebox{-4mm}{\hspace{-3mm}\textcolor{myPurple}{D}}}
\psfrag{A5}[c][c][.70]{\raisebox{3mm}{\hspace{0mm}\textcolor{myGreen}{E}}}
\psfrag{A6}[c][c][.70]{\raisebox{0mm}{\hspace{2mm}\textcolor{myBlue}{F}}}

\psfrag{xt0}[c][c][.70]{\raisebox{-3.1mm}{\hspace{0mm}$0$}}
\psfrag{xt50}[c][c][.70]{\raisebox{-3.1mm}{\hspace{0mm}$50$}}
\psfrag{xt100}[c][c][.70]{\raisebox{-3.1mm}{\hspace{0mm}$100$}}
\psfrag{xt150}[c][c][.70]{\raisebox{-3.1mm}{\hspace{0mm}$150$}}
\psfrag{xt200}[c][c][.70]{\raisebox{-3.1mm}{\hspace{0mm}$200$}}
\psfrag{xtm50}[c][c][.70]{\raisebox{-3.5mm}{\hspace{0mm}$-50$}}
\psfrag{xtm100}[c][c][.70]{\raisebox{-3.5mm}{\hspace{0mm}$-100$}}
\psfrag{xtm150}[c][c][.70]{\raisebox{-3.5mm}{\hspace{0mm}$-150$}}
\psfrag{xtm200}[c][c][.70]{\raisebox{-3.5mm}{\hspace{0mm}$-200$}}

\psfrag{yt0}[r][r][.70]{\raisebox{0mm}{\hspace{0mm}$0$}}
\psfrag{yt50}[r][r][.70]{\raisebox{0mm}{\hspace{0mm}$50$}}
\psfrag{yt100}[r][r][.70]{\raisebox{0mm}{\hspace{0mm}$100$}}
\psfrag{yt150}[r][r][.70]{\raisebox{0mm}{\hspace{0mm}$150$}}
\psfrag{ytm50}[r][r][.70]{\raisebox{0mm}{\hspace{0mm}$-50$}}
\psfrag{ytm100}[r][r][.70]{\raisebox{0mm}{\hspace{0mm}$-100$}}
\psfrag{ytm150}[r][r][.70]{\raisebox{0mm}{\hspace{0mm}$-150$}}

\includegraphics[width=.65\columnwidth, keepaspectratio]{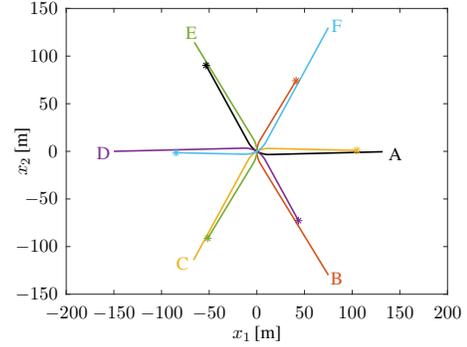}

\caption{Trajectories of the six targets. The stars mark the final target positions.}
\label{fig:simulated_scenario}

\vspace{-3mm}

\end{figure}

We simulated 
six targets ``A''
through ``F'' that move in a rectangular
region of interest (ROI)
during 140 time steps with a constant speed of 1 m/s.
The target trajectories and the ROI are shown in Fig.~\ref{fig:simulated_scenario}.
The targets move toward the ROI center starting from positions uniformly placed on a circle of radius 150 m,
and then, approximately at time $n \! = \! 75$, they
perform a right turn of 60 degrees.
Targets
A, C, and E start and stop to exist at times $n\! = \!10$ and $n \! = \! 130$, respectively, and
targets 
B, D, and F 
at times $n\!=1\!$ and $n\!=\! 120$, respectively. There are $C \! = \! 3$ target classes;
targets
A and D belong to class $c \! = \! 1$, targets
B and E 
to class  $c \! = \! 2$, and targets
C and F
to class $c \! = \! 3$.
However, the tracking method has no prior knowledge about these
class affiliations.

The PT states are $\V{x}_{n,k} \!\rmv =\! [\check{\V{x}}_{n,k}^{\T},\dot{\check{\V{x}}}_{n,k}^{\T}]^{\T} \rmv$
with two-dimensional (2D) position $\check{\V{x}}_{n,k}$ and 2D velocity $\dot{\check{\V{x}}}_{n,k}$.
The dynamic model used by the tracking methods
is a nearly constant velocity model, 
i.e., $\V{x}_{n,k} \! = \! \V{\theta} (\V{x}_{n-1,k},\V{u}_{n,k}) \! = \! \V{A} \V{x}_{n-1,k} \rmv + \V{W} \rmv \V{u}_{n,k}$,
where 
$\V{A} \rmv\in \mathbb{R}^{4 \times 4}$ and $\V{W} \!\in\rmv \mathbb{R}^{4 \times 2}$ are 
as in \cite[Sec.\ 6.3.2]{BarRonKir:01} 
(with a time step duration of 2 s) and 
$\V{u}_{n,k}$ is iid zero-mean Gaussian with a 
per-component 
standard deviation of 0.1 m/s$^2$.
Note that the PTs' dynamic model does not depend on the PT class, 
but the 
PT class 
may be related to some other target characteristic(s) 
such as color, size, or type.
In the tracking methods,
the number of PTs is chosen as $K \! = \! 20$.
The elements of the class transition matrix
\rd{$\V{D}(\V{x}_{n-1,k})\! = \! \V{D}$}
are chosen as
$\rd{[\V{D}]_{i,j}} \! =\rmv 0.95$ if $i\!=\!j$ and
$\rd{[\V{D}]_{i,j}} \! = \rmv 0.025$ if $i \! \neq \! j$.

There are 
$S \! = \! 1$ or $2$ sensors
equally spaced on a 
circle of radius 3 km around
$(0,0)$.
The sensors 
measure 
range and bearing.
The target-generated
measurements are thus modeled 
as 
$\V{q}_{n,m}^{(s)} \! = \rmv \V{\psi}_{s} \big( \V{x}_{n,k},\V{v}_{n,m}^{(s)} \big) \! = \! \big[ \| \check{\V{x}}_{n,k} - \V{p}^{(s)} \| ,\phi(\check{\V{x}}_{n,k},\V{p}^{(s)}) \big]^{\T}$\linebreak $+\, \V{v}_{n,m}^{(s)}$, 
where $\V{p}^{(s)}\rmv$
is the position of sensor $s$,
$\phi( \check{\V{x}}_{n,k}, \V{p}^{(s)} )$ is the 
angle 
of the vector $\check{\V{x}}_{n,k} - \V{p}^{(s)}\rmv$, and $\V{v}_{n,m}^{(s)}$ 
is iid---also across $s$---zero-mean Gaussian with covariance matrix 
$\text{diag}(\sigma_\text{r}^2, \sigma_\text{b}^2)$,
where $\sigma_\text{r} \! = \! 5$ m and $\sigma_\text{b} \! = \! 0.1^\circ\rmv$. 
The false alarm pdf $f_{0}(\V{q}_{n,m}^{(s)})$ is linearly increasing with respect to 
range and uniform with respect to bearing
within the ROI,
and zero outside the ROI. The mean number of false alarms, $\mu^{(s)}\rmv$, is 5, 10, or 20. The probability of detection is 
$P_{\text{d}}^{(s)}(\V{x}_{n,k},\ell_{n,k}) \rmv = \! P_{\text{d}}$\linebreak $= \rmv 0.9$.
Each 
sensor includes a classifier whose output $\zeta_{n,m}^{(s)} \! \in \! \{ 0,1,2,3 \}$ accompanies measurement $\V{q}_{n,m}^{(s)}$, where 
$\zeta_{n,m}^{(s)} \! = \! 0$ expresses the classifier's
belief that $\V{q}_{n,m}^{(s)}$
is clutter-generated and $\zeta_{n,m}^{(s)} \! = \! 1$, $2$, and $3$
that it is generated by a target belonging to class $1$, $2$, and $3$, respectively.
The elements 
of the classifiers' confusion matrix
$\V{G}^{(s)}(\V{x}_{n,k}) \! = \! \V{G}$
are chosen 
as
$\rd{[\V{G}]_{i,j}}  \!= \! 0.85$ if $i \!= \! j$ and
$\rd{[\V{G}]_{i,j}} \!= \! 0.05$ if $i \! \neq \! j$, with
$i \! \in \! \{ 0,1,2,3 \}$, $j \! \in\! \{ 1,2,3 \}$. The pmf $p_0\big(\zeta_{n,m}^{(s)} \big| \V{q}_{n,m}^{(s)}\big)$
is chosen  independently of $\V{q}_{n,m}^{(s)}$ as $p_0\big(\zeta_{n,m}^{(s)}\!=\!0\big)\!= \! 0.85$ and $p_0\big(\zeta_{n,m}^{(s)}\!=\!i\big)\!= \! 0.05$ for $i \! \in \! \{1,2,3\}$.

\vspace{-2mm}

\subsection{Results}\label{subsec:results}

\vspace{-.3mm}

We compare the performance of the proposed classifier-aided method with that of the baseline method of \cite{MeyBraWilHla:J17}, which 
does not use the classifier output $\zeta_{n,m}^{(s)}$.
The performance is assessed in terms of the time-averaged mean generalized optimal sub-pattern assignment (MGOSPA) error \cite{RahGarSven:C17},
the time-averaged mean optimal sub-pattern assignment (MOSPA) error \cite{SchVoVo:J08},
the time-averaged MOSPA-for-tracks (MOSPA-T) error~\cite{RisVoClarkVo:J11}
(all three with order 1 and cutoff parameter 20 m; the time-averaged MOSPA-T error additionally with label error penalty 20 m),
and the false alarm rate (FAR).
The MGOSPA and MOSPA errors
take into account estimation errors for correctly detected targets and errors due to incorrect target detections. The MOSPA-T error
additionally penalizes incorrect switches of the estimated tracks; it equals the MOSPA error when there are no switches.
The FAR is the number of false tracks per unit of space and unit of time.

\begin{figure}[!t]

\centering

\psfrag{basT}[c][c][.65]{\raisebox{2mm}{\hspace{6mm}Baseline \cite{MeyBraWilHla:J17}}}
\psfrag{proT}[c][c][.65]{\raisebox{2mm}{\hspace{6mm}Proposed}}

\psfrag{bas1}[l][l][.65]{\hspace{.5mm}$S \! = \! 1$}
\psfrag{bas2}[l][l][.65]{\hspace{.5mm}$S \! = \! 2$}
\psfrag{pro1}[l][l][.65]{\hspace{.5mm}$S \! = \! 1$}
\psfrag{pro2}[l][l][.65]{\hspace{.5mm}$S \! = \! 2$}

\psfrag{xt0}[c][c][.70]{\raisebox{-3mm}{\hspace{0mm}$0$}}
\psfrag{xt20}[c][c][.70]{\raisebox{-3mm}{\hspace{0mm}$20$}}
\psfrag{xt40}[c][c][.70]{\raisebox{-3mm}{\hspace{0mm}$40$}}
\psfrag{xt60}[c][c][.70]{\raisebox{-3mm}{\hspace{0mm}$60$}}
\psfrag{xt80}[c][c][.70]{\raisebox{-3mm}{\hspace{0mm}$80$}}
\psfrag{xt100}[c][c][.70]{\raisebox{-3mm}{\hspace{0mm}$100$}}
\psfrag{xt120}[c][c][.70]{\raisebox{-3mm}{\hspace{0mm}$120$}}
\psfrag{xt140}[c][c][.70]{\raisebox{-3mm}{\hspace{0mm}$140$}}

\psfrag{yt0}[r][r][.70]{\raisebox{0mm}{\hspace{0mm}$0$}}
\psfrag{yt5}[r][r][.70]{\raisebox{0mm}{\hspace{0mm}$5$}}
\psfrag{yt10}[r][r][.70]{\raisebox{0mm}{\hspace{0mm}$10$}}
\psfrag{yt15}[r][r][.70]{\raisebox{0mm}{\hspace{0mm}$15$}}
\psfrag{yt20}[r][r][.70]{\raisebox{0mm}{\hspace{0mm}$20$}}
\psfrag{yt25}[r][r][.70]{\raisebox{0mm}{\hspace{0mm}$25$}}

\psfrag{xla}[c][c][.75]{\raisebox{-7.5mm}{Time Step $n$}}
\psfrag{yla}[c][c][.75]{\raisebox{7mm}{MOSPA-T Error [m]}}

{\hspace*{2mm}\includegraphics[width=.84\columnwidth, height=3.8cm]{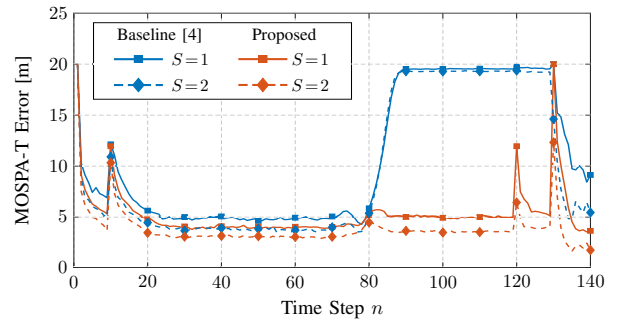}}

\vspace{1.2mm}

\caption{MOSPA-T error versus time for $\mu^{(s)} \! = \! 10$.}
\label{fig:MOSPA_comp}

\vspace{-3mm}

\end{figure}

Tables~\ref{tab:MOSPA_ToT_FAR_s1} and \ref{tab:MOSPA_ToT_FAR_s2} report these metrics, averaged over 200 simulation runs, for $S \! = \! 1$ and $S \! = \! 2$ \rd{sensors}, 
respectively, and for three different values of $\mu^{(s)}\rmv$.
It can be seen that the proposed method 
consistently outperforms the baseline method; the gain in performance increases with the clutter level. 
In particular, for $\mu^{(s)} \! = \! 20$, the FAR is about 4.5 times lower
than that obtained with the baseline method,
and the time-averaged  MOSPA-T error 
is about 53\,\% (for $S \! = \! 1$) and 62\,\% (for $S \! = \! 2$)\linebreak lower.
Fig.~\ref{fig:MOSPA_comp} displays
the MOSPA-T error 
versus time, for $S \! = \! 1$ or $2$ sensors and mean number of false alarms 
$\mu^{(s)} \!=\!10$. Again, the proposed method
consistently outperforms\linebreak the baseline method.
In particular, the MOSPA-T error of the baseline method noticeably increases right after the turn of the targets (approximately at time $n \! = \! 78$), which indicates
the occurrence of switches
among the estimated tracks. The MOSPA-T error of the proposed method, instead, keeps approximately
the same level,
which indicates that no or almost no track switches occur.
Note that the
peaks observed at various times
are due to target appearance and disappearance.

\begin{table}[!t]
\newcolumntype{Y}[1]{>{\hsize=#1\hsize\centering\arraybackslash}X}
\vspace{0mm}

\renewcommand{\arraystretch}{1.3}
\scriptsize

\centering

\caption{\hspace{5mm}time-averaged mgospa, mospa, and mospa-t errors \newline as well as far for $S \! = \! 1$ \rd{sensor}.}

\vspace{-1mm}

\begin{tabularx}{\columnwidth}{ Y{1} | Y{1} Y{1} | Y{1} Y{1} | Y{1} Y{1} | Y{1} Y{1} }
\hline
\multirow{3}{*}{$\mu^{(s)}$}	&	\multicolumn{2}{c |}{\multirow{2}{*}{MGOSPA [m]}}	&	\multicolumn{2}{c |}{\multirow{2}{*}{MOSPA [m]}}	&	\multicolumn{2}{c |}{\multirow{2}{*}{MOSPA-T [m]}}	&	\multicolumn{2}{c }{FAR}		\\
	&	&	&	&	&	&	&	\multicolumn{2}{c }{[$\text{km}^{-2} \text{s}^{-1}$]}		\\
		&	Basel.	&	Prop.	&	Basel.	&	Prop.	&	Basel.	&	Prop.	&	Basel.	&	Prop.	\\	\hline \hline
$5$		&	$22.4$	&	$20.8$	&	$5.0$	&	$4.4$	&	$10.0$	&	$4.7$	&	$0.64$	&	$0.31$	\\[-.5mm]
$10$	&	$25.8$	&	$21.9$	&	$6.0$	&	$4.8$	&	$10.7$	&	$5.2$	&	$1.38$	&	$0.53$	\\[-.5mm]
$20$	&	$39.1$	&	$24.2$	&	$8.6$	&	$5.6$	&	$12.7$	&	$6.0$	&	$4.62$	&	$1.01$	\\ \hline

\end{tabularx}

\label{tab:MOSPA_ToT_FAR_s1}	
	
\vspace{-1mm}	
	
\end{table}

\begin{table}[!t]
\newcolumntype{Y}[1]{>{\hsize=#1\hsize\centering\arraybackslash}X}
\vspace{0mm}

\renewcommand{\arraystretch}{1.3}
\scriptsize

\centering

\caption{\hspace{5mm}time-averaged mgospa, mospa, and mospa-t errors \newline as well as far for $S \! = \! 2$ \rd{sensors}.}

\vspace{-1mm}

\begin{tabularx}{\columnwidth}{ Y{1} | Y{1} Y{1} | Y{1} Y{1} | Y{1} Y{1} | Y{1} Y{1} }
\hline
\multirow{3}{*}{$\mu^{(s)}$}	&	\multicolumn{2}{c |}{\multirow{2}{*}{MGOSPA [m]}}	&	\multicolumn{2}{c |}{\multirow{2}{*}{MOSPA [m]}}	&	\multicolumn{2}{c |}{\multirow{2}{*}{MOSPA-T [m]}}	&	\multicolumn{2}{c }{FAR}		\\
	&	&	&	&	&	&	&	\multicolumn{2}{c }{[$\text{km}^{-2} \text{s}^{-1}$]}		\\
		&	Basel.	&	Prop.	&	Basel.	&	Prop.	&	Basel.	&	Prop.	&	Basel.	&	Prop.	\\	\hline \hline
$5$		&	$16.7$	&	$15.7$	&	$3.5$	&	$3.3$	&	$9.0$	&	$3.5$	&	$0.37$	&	$0.21$	\\[-.5mm]
$10$	&	$19.5$	&	$16.4$	&	$4.4$	&	$3.5$	&	$9.6$	&	$3.8$	&	$1.05$	&	$0.36$	\\[-.5mm]
$20$	&	$30.0$	&	$18.4$	&	$7.1$	&	$4.2$	&	$11.6$	&	$4.5$	&	$3.69$	&	$0.83$	\\ \hline

\end{tabularx}

\label{tab:MOSPA_ToT_FAR_s2}	
	
\vspace{1mm}	
	
\end{table}

These results
demonstrate the benefit of using class information 
to improve 
the performance of the
SPA-based MTT 
methodology.
Indeed, the class information allows a more reliable association between measurements and targets, which reduces track switches and thus increases 
tracking accuracy.
Additional results assessing the performance of the proposed method for different values of the number of classes $C$ and different choices
of the confusion matrix $\V{G}$ are provided in the supplementary material manuscript \cite{SupplMat}.

\vspace{-2mm}

\section{Conclusion}
\label{sec:conclusion}
A challenging issue in MTT is 
the unknown association between measurements and targets.
The use of 
class information in addition to sensor measurements
can improve 
probabilistic target-measurement association and,
in turn, 
overall MTT
performance.
In this letter, we 
showed how the
output of a classifier
can be 
integrated 
into the SPA-based MTT framework recently proposed in \cite{MeyBraWilHla:J17}, 
and we demonstrated 
experimentally 
significant performance 
advantages of the resulting classifier-aided method 
over the 
method of \cite{MeyBraWilHla:J17}.

\renewcommand{\baselinestretch}{0.97}
\selectfont
\bibliographystyle{IEEEtran}
\bibliography{../../Bibliography/IEEEabrv,../../Bibliography/myBib,../../Bibliography/Wgroup,../../Bibliography/BiblioCV,../../Bibliography/Temp,../../Bibliography/biblio}

\end{document}


\onehalfspacing

\title{{\bf Classification-Aided Multitarget Tracking Using the\\[-1mm]Sum-Product Algorithm --- Supplementary Material}\vspace{2mm}}

\author{
		Domenico Gaglione\thanks{Co-first authors.}\ist\,, 		Giovanni Soldi$^{*}\!$, 
		Paolo Braca, 		 Giovanni De Magistris, 
 		Florian Meyer,\\
		and Franz Hlawatsch\vspace{2mm}
}\date{\today\vspace{0mm}}

\maketitle

\setcounter{tocdepth}{2}

\renewcommand{\baselinestretch}{1.13}\small\normalsize

\noindent This\blfootnote{
This work was supported in part by the NATO Allied Command Transformation (ACT) under the DKOE project
and by the Austrian Science Fund (FWF) under grants J 3886-N31 and 
P 32055-N31.
D.\ Gaglione, G.\ Soldi, P.\ Braca, and G.\ De\ Magistris are with the NATO Centre for Maritime Research and Experimentation (CMRE), La~Spezia, Italy (e-mail: [domenico.gaglione, giovanni.soldi, paolo.braca, giovanni.demagistris]@cmre.nato.int). 
F.\ Meyer 
is with the Scripps Institution of Oceanography and the Electrical and Computer Engineering Department, University of California San Diego, La Jolla, CA, USA (e-mail: flmeyer@ucsd.edu).
F.\ Hlawatsch is with the Institute of Telecommunications, TU Wien, Vienna, Austria (e-mail: franz.hlawatsch@tuwien.ac.at).}
manuscript supplements the related manuscript, `Classification-Aided Multitarget Tracking Using the Sum-Product Algorithm' \cite{GagSolBraMagMeyHla:J19}
by the same authors. The presented material comprises a Bayesian network related to the transition probability density function (pdf) for the augmented state
in \cite[Eq.~(1)]{GagSolBraMagMeyHla:J19}; 
a Bayesian network related to the likelihood function in
\cite[Eq.~(2)]{GagSolBraMagMeyHla:J19};
a Bayesian network related to the joint pdf $f \big( \V{y}, \V{a}, \V{b}, \V{m}, \V{z} \big)$;
a derivation of the factorization of the joint posterior pdf in \cite[Eq.~(3)]{GagSolBraMagMeyHla:J19};
a detailed statement and description of the multitarget tracking method proposed in \cite[Sec. III]{GagSolBraMagMeyHla:J19}; 
and additional simulation results.
Basic definitions, notation, and assumptions are given in \cite{GagSolBraMagMeyHla:J19} and will be repeated here only partly.

\section{Bayesian Networks 
Related to the Augmented State Transition pdf\\$f(\V{y}_{n,k}|\V{y}_{n-1,k})$ and the Likelihood Function $f\big(\V{z}_{n,m}^{(s)} \big| \V{x}_{n,k},\ell_{n,k} \big)$}

\begin{figure}[!t]
\vspace{0mm}
\renewcommand{\arraystretch}{1.2}
\scriptsize
\begin{subfigure}[t]{.48\linewidth}
\centering
\psfrag{aa}[c][c][1.2]{\raisebox{0mm}{\hspace{0mm}$\bar{\V{x}}_{k}$}}
\psfrag{rk}[c][c][1.2]{\raisebox{0mm}{\hspace{0mm}$\bar{r}_{k}$}}
\psfrag{cc}[c][c][1.2]{\raisebox{0mm}{\hspace{0mm}$\bar{\ell}_{k}$}}
\psfrag{dd}[c][c][1.2]{\raisebox{0mm}{\hspace{0mm}$\V{x}_{k}$}}
\psfrag{ee}[c][c][1.2]{\raisebox{0mm}{\hspace{0mm}$r_{k}$}}
\psfrag{ff}[c][c][1.2]{\raisebox{0mm}{\hspace{0mm}$\ell_{k}$}}
\includegraphics[scale=1.2]{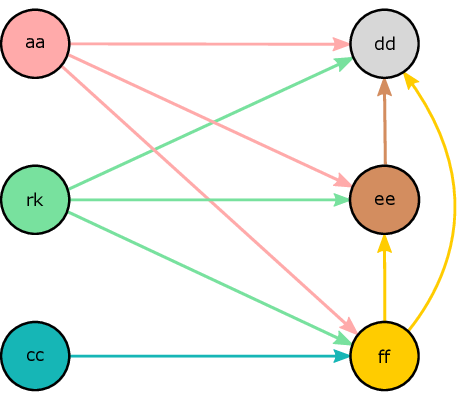}
\caption{}
\label{fig-S:bayesian-network_transitionPdf}		
\end{subfigure}
\hfill
\begin{subfigure}[t]{.48\linewidth}
\centering
\psfrag{xk}[c][c][1.2]{\raisebox{0mm}{\hspace{0mm}$\V{x}_{k}$}}
\psfrag{lk}[c][c][1.2]{\raisebox{0mm}{\hspace{0mm}$\ell_{k}$}}
\psfrag{cm}[c][c][1.2]{\raisebox{0mm}{\hspace{0mm}$\zeta_{m}$}}
\psfrag{qm}[c][c][1.2]{\raisebox{0mm}{\hspace{0mm}$\V{q}_{m}$}}
\includegraphics[scale=1.2]{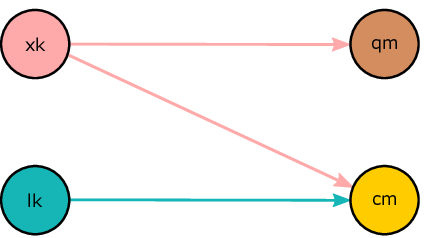}
\caption{}
\label{fig-S:bayesian-network_likelihood}
\end{subfigure}
\vspace{-1mm}
\caption{\small{Bayesian networks related to (a) the factorization of the augmented state transition pdf 
$f(\V{y}_{n,k}|\V{y}_{n-1,k}) \!=$ $f(\V{x}_{n,k},r_{n,k},\ell_{n,k} | \V{x}_{n-1,k},r_{n-1,k},\ell_{n-1,k})$ in \eqref{eq-S:augmented_state_transition_pdf} and 
(b) the factorization of the likelihood function $f\big(\V{z}_{n,m}^{(s)} \big| \V{x}_{n,k},\ell_{n,k} \big) \!=\! f\big( \V{q}_{n,m}^{(s)}, \zeta_{n,m}^{(s)} \big| \V{x}_{n,k},\ell_{n,k} \big)$ 
in \eqref{eq:factorisation_likelihood_ais_single_cluster_1}.
For simplicity, the following short notations are used: $\protect\bar{\V{x}}_{k} \! \deq \rmv \V{x}_{n-1,k}$, $\protect\bar{r}_k \! \deq \rmv r_{n-1,k}$, 
$\protect\bar{\ell}_k \! \deq$ $\ell_{n-1,k}$, $\V{x}_{k} \! \deq \rmv \V{x}_{n,k}$, $r_k \! \deq \rmv r_{n,k}$, $\ell_k \! \deq \rmv \ell_{n,k}$, 
$\V{q}_m \! \deq \rmv \V{q}_{n,m}^{(s)}$, and $\zeta_{m} \! \deq \rmv \zeta_{n,m}^{(s)}$. The color of each edge matches the color of the associated parent node.}}	
\label{fig-S:bayesian-network}
\vspace{3mm}
\end{figure}

Let us consider the augmented state transition pdf $f(\V{y}_{n,k}|\V{y}_{n-1,k}) = f(\V{x}_{n,k},r_{n,k},\ell_{n,k} | \V{x}_{n-1,k},r_{n-1,k},\ell_{n-1,k})$.
Starting out from the factorization in \cite[Eq.~(1)]{GagSolBraMagMeyHla:J19},
we can further factorize $f(\V{x}_{n,k},r_{n,k},\ell_{n,k} | \V{x}_{n-1,k},r_{n-1,k}, \linebreak \ell_{n-1,k})$ according to
\begin{align}
&f(\V{x}_{n,k},r_{n,k},\ell_{n,k} | \V{x}_{n-1,k},r_{n-1,k},\ell_{n-1,k}) \nn\\[0mm]
&\quad= f(\V{x}_{n,k} | r_{n,k}, \ell_{n,k}, \V{x}_{n-1,k},r_{n-1,k}) \, p (r_{n,k} | \ell_{n,k},\V{x}_{n-1,k},r_{n-1,k}) \ist p (\ell_{n,k} |\V{x}_{n-1,k},r_{n-1,k},\ell_{n-1,k}) \ist.
\label{eq-S:augmented_state_transition_pdf} 
\end{align}
The Bayesian network illustrating the statistical dependencies of the random variables involved in this expression is shown in Fig.~\ref{fig-S:bayesian-network_transitionPdf}.

Next, we consider the likelihood function $f\big(\V{z}_{n,m}^{(s)} \big| \V{x}_{n,k},\ell_{n,k} \big) = f\big( \V{q}_{n,m}^{(s)}, \zeta_{n,m}^{(s)} \big| \V{x}_{n,k},\ell_{n,k} \big)$. 
According to \cite[Eq.~(2)]{GagSolBraMagMeyHla:J19}, we have
\be
f\big( \V{q}_{n,m}^{(s)}, \zeta_{n,m}^{(s)} \big| \V{x}_{n,k}, \ell_{n,k} \big) = p \big( \zeta_{n,m}^{(s)} \big| \V{x}_{n,k}, \ell_{n,k} \big)  \ist f \big( \V{q}_{n,m}^{(s)} \big| \V{x}_{n,k} \big).
\label{eq:factorisation_likelihood_ais_single_cluster_1}
\ee
The Bayesian network illustrating the statistical dependencies of the involved random variables is shown in Fig.~\ref{fig-S:bayesian-network_likelihood}.

\section{Factorization of the Joint pdf $f(\V{y},\V{a},\V{b}, \V{m}, \V{z})$ and Related Bayesian\\Network}
\label{sec:Bay_Net_Der_Joint_Post}

In this section, we develop a factorization of the joint
pdf $f(\V{y},\V{a}, \V{b}, \V{m}, \V{z})$
and we show the related Bayesian network.
Let us recall some relevant notation: we define
$\V{y}_{n} \!\triangleq\rmv [\V{y}^\T_{n,1}, \ldots, \V{y}^\T_{n,K}]^\T\rmv$, 
$\V{y} \rmv\triangleq\rmv [\V{y}^\T_{0},  \ldots, \V{y}^\T_{n}]^\T\rmv$,
$\V{z}^{(s)}_n \!\triangleq \big[\V{z}^{(s)\T}_{n,1}, \ldots, \V{z}^{(s)\T}_{n,M^{(s)}_n} \big]^{\T} \rmv$,
$\V{z}_n \!\triangleq\rmv \big[\V{z}^{(1)\T}_n, \ldots, \V{z}^{(S)\T}_n \big]^{\T}\rmv$,
$\V{z} \rmv\triangleq\rmv [\V{z}^\T_1, \ldots, \V{z}^\T_n]^{\T}\rmv$,
$\V{a}^{(s)}_{n} \!\triangleq\rmv \big[a^{(s)}_{n,1},\ldots,a^{(s)}_{n,K} \big]^{\T}\rmv$,
$\V{a}_n \!\triangleq\rmv\big[\V{a}^{(1)\T}_n \rmv, \ldots, \linebreak \V{a}^{(S)\T}_n \big]^{\T}\rmv$, 
$\V{a} \rmv\triangleq\rmv [\V{a}_1^\T, \ldots, \V{a}_n^\T]^\T\rmv$,
$\V{b}_{n}^{(s)} \!\triangleq\rmv \big[b_{n,1}^{(s)}, \ldots, b_{n,M_n^{(s)}}^{(s)} \big]^{\T}\rmv$,
$\V{b}_n \!\triangleq\rmv \big[\V{b}^{(1)\T}_n \rmv, \ldots, \V{b}^{(S)\T}_n \big]^{\T}\rmv$, 
$\V{b} \rmv\triangleq\rmv \big[\V{b}_1^{\T}, \ldots, \V{b}^{\T}_n \big]^{\T}\rmv$, 
$\V{m}_n \!\triangleq \big[M^{(1)}_n, \ldots, M^{(S)}_n \big]^{\T}\rmv$, and $\V{m} \rmv\triangleq\rmv [\V{m}^\T_1,  \ldots, \V{m}^\T_n]^{\T}\rmv$.
Our derivation 
is based on the following commonly used assumptions:

\begin{itemize}

\item[A1)] The augmented state vector $\V{y}_{n}$ evolves over time $n$ according to a first-order Markov model \cite{Mah:B07,BarWilTia:B11}.

\item[A2)] The augmented states $\V{y}_{n,k}$ of different potential targets (PTs) $k$ evolve independently \cite{MeyBraWilHla:J17,MeyKroWilLauHlaBraWin:J18,SoldiMBH:J19}.

\item[A3)] Given $\V{y}_{n}$,
the association vectors $\V{a}^{(s)}_n$ and $\V{b}^{(s)}_n$ and the number of measurements $M_n^{(s)}$ are
conditionally independent of $\V{y}_{n'}$ with $n' \! \neq \! n$, and
conditionally independent of $\V{a}_{n'}^{(s')}\rmv$, $\V{b}_{n'}^{(s')}\rmv$, and $M_{n'}^{(s')}$ with $n' \! \neq \! n$ and $s' \! \in \! \{ 1, \ldots, S \}$ or with $n' \! = \! n$ and $s' \! \in \! \{ 1, \ldots, S \} \setminus \{ s \}$ \cite{MeyBraWilHla:J17,MeyKroWilLauHlaBraWin:J18,SoldiMBH:J19}.

\item[A4)] Given $\V{y}$, $\V{a}$, and $\V{m}$, the measurement vector $\V{z}$ is conditionally independent of $\V{b}$ \cite{MeyBraWilHla:J17,MeyKroWilLauHlaBraWin:J18,SoldiMBH:J19}. That is, $f ( \V{z} | \V{y}, \V{a}, \linebreak \V{b}, \V{m}  ) \! = \! f ( \V{z} | \V{y}, \V{a}, \V{m} )$.

\item[A5)] Given $\V{y}$,  $\V{a}$, and $\V{m}$, the measurement vectors $\V{z}_n^{(s)}$ are conditionally independent across 
$n$ and 
$s$ \cite{MeyBraWilHla:J17,MeyKroWilLauHlaBraWin:J18,SoldiMBH:J19}.

\item[A6)]
Given $\V{y}_{n}$,  $\V{a}_{n}^{(s)}\rmv$, and $M_{n}^{(s)}\rmv$, the measurement vector $\V{z}_n^{(s)}$ is conditionally independent of $\V{y}_{n'}$ with $n' \! \neq \! n$, and conditionally independent of $\V{a}_{n'}^{(s')}$ and $M_{n'}^{(s')}$ with $n' \! \neq \! n$ and $s' \! \in \! \{ 1, \ldots, S \}$ or with $n' \! = \! n$ and $s' \! \in \! \{ 1, \ldots, S \} \setminus \{ s \}$ \cite{MeyBraWilHla:J17,MeyKroWilLauHlaBraWin:J18,SoldiMBH:J19}.
That is, $f\big( \V{z}_{n}^{(s)} \big| \V{y}, \V{a}, \V{m} \big) \! = \! f\big( \V{z}_{n}^{(s)} \big| \V{y}_{n}, \V{a}_{n}^{(s)}, M_{n}^{(s)} \big)$.

\item[A7)] Given $\V{y}_{n}$, $\V{a}^{(s)}_{n}\rmv$, and $M^{(s)}_{n}\rmv$, the measurement vectors $\V{z}^{(s)}_{n,m}\ist$
are conditionally independent across $m$ \cite{MeyBraWilHla:J17,MeyKroWilLauHlaBraWin:J18,SoldiMBH:J19}.

\item[A8)] Given $\V{a}_{n}^{(s)}$ and $M_{n}^{(s)}\rmv$, the association vector $\V{b}_{n}^{(s)}$ is conditionally independent of $\V{y}_n$ \cite{MeyBraWilHla:J17,MeyKroWilLauHlaBraWin:J18,SoldiMBH:J19}.
That is, $p\big(\V{b}_{n}^{(s)} \big| \V{a}_{n}^{(s)}\rmv, M_{n}^{(s)}\rmv, \V{y}_{n}\big) \! = \! p\big(\V{b}_{n}^{(s)} \big| \V{a}_{n}^{(s)}\rmv, M_{n}^{(s)}\big)$.

\end{itemize}

\begin{figure}[t!]
\centering

\psfrag{timep}[c][c][.85]{\raisebox{0mm}{\hspace{0mm}$n-1$}}
\psfrag{timec}[c][c][.85]{\raisebox{0mm}{\hspace{0mm}$n$}}

\psfrag{yp1}[c][c][.85]{\raisebox{1mm}{\hspace{0mm}$\bar{\V{y}}_{1}$}}
\psfrag{ypk}[c][c][.85]{\raisebox{1mm}{\hspace{0mm}$\bar{\V{y}}_{K}$}}

\psfrag{bp1}[c][c][.85]{\raisebox{1mm}{\hspace{0mm}$\bar{\V{b}}^{1}$}}
\psfrag{ap1}[c][c][.85]{\raisebox{1mm}{\hspace{0mm}$\bar{\V{a}}^{1}$}}
\psfrag{mp1}[c][c][.85]{\raisebox{1mm}{\hspace{-.5mm}$\bar{M}^{1}$}}
\psfrag{zp11}[c][c][.85]{\raisebox{1mm}{\hspace{0mm}$\bar{\V{z}}^{1}_{1}$}}
\psfrag{zp1m}[c][c][.85]{\raisebox{1mm}{\hspace{0mm}$\bar{\V{z}}^{1}_{\bar{M}^1}$}}

\psfrag{bp2}[c][c][.85]{\raisebox{1mm}{\hspace{0mm}$\bar{\V{b}}^{2}$}}
\psfrag{ap2}[c][c][.85]{\raisebox{1mm}{\hspace{0mm}$\bar{\V{a}}^{2}$}}
\psfrag{mp2}[c][c][.85]{\raisebox{1mm}{\hspace{-.5mm}$\bar{M}^{2}$}}
\psfrag{zp21}[c][c][.85]{\raisebox{1mm}{\hspace{0mm}$\bar{\V{z}}^{2}_{1}$}}
\psfrag{zp2m}[c][c][.85]{\raisebox{1mm}{\hspace{0mm}$\bar{\V{z}}^{2}_{\bar{M}^2}$}}

\psfrag{yc1}[c][c][.85]{\raisebox{1mm}{\hspace{0mm}${\V{y}}_{1}$}}
\psfrag{yck}[c][c][.85]{\raisebox{1mm}{\hspace{0mm}${\V{y}}_{K}$}}

\psfrag{bc1}[c][c][.85]{\raisebox{1mm}{\hspace{0mm}$\V{b}^{1}$}}
\psfrag{ac1}[c][c][.85]{\raisebox{1mm}{\hspace{0mm}${\V{a}}^{1}$}}
\psfrag{mc1}[c][c][.85]{\raisebox{1mm}{\hspace{-.5mm}${M}^{1}$}}
\psfrag{zc11}[c][c][.85]{\raisebox{1mm}{\hspace{0mm}${\V{z}}^{1}_{1}$}}
\psfrag{zc1m}[c][c][.85]{\raisebox{1mm}{\hspace{0mm}${\V{z}}^{1}_{{M}^1}$}}

\psfrag{bc2}[c][c][.85]{\raisebox{1mm}{\hspace{0mm}$\V{b}^{2}$}}
\psfrag{ac2}[c][c][.85]{\raisebox{1mm}{\hspace{0mm}${\V{a}}^{2}$}}
\psfrag{mc2}[c][c][.85]{\raisebox{1mm}{\hspace{-.5mm}${M}^{2}$}}
\psfrag{zc21}[c][c][.85]{\raisebox{1mm}{\hspace{0mm}${\V{z}}^{2}_{1}$}}
\psfrag{zc2m}[c][c][.85]{\raisebox{1mm}{\hspace{0mm}${\V{z}}^{2}_{{M}^2}$}}

\psfrag{sm1}[c][c][.85]{\raisebox{1mm}{\hspace{0mm}$s=1$}}
\psfrag{sm2}[c][c][.85]{\raisebox{1mm}{\hspace{-.5mm}$s=2$}}
\psfrag{s1}[c][c][.85]{\raisebox{1mm}{\hspace{0mm}$s=1$}}
\psfrag{s2}[c][c][.85]{\raisebox{1mm}{\hspace{0mm}$s=2$}}

\includegraphics[scale=1.0]{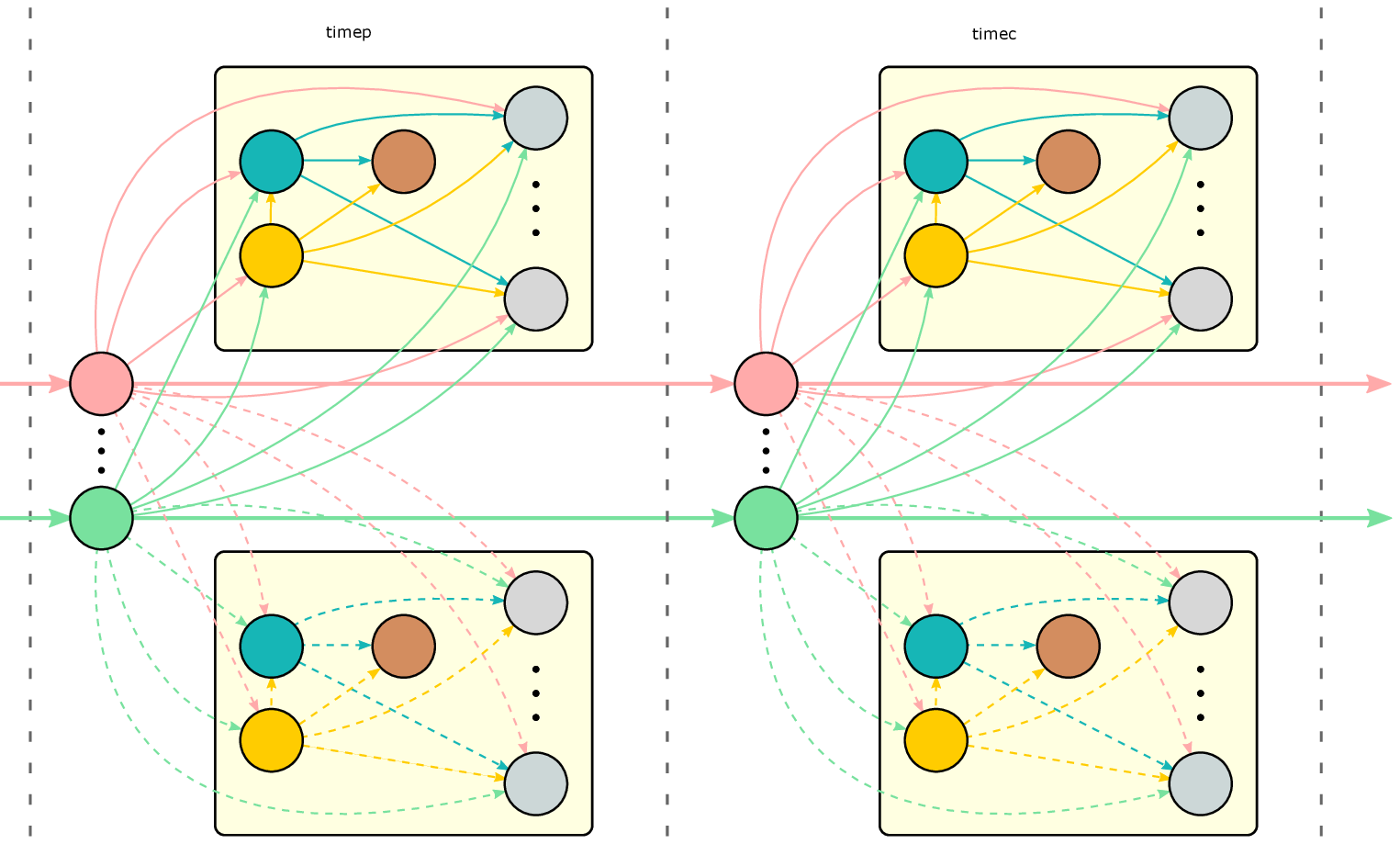}
\vspace{1.7mm}
\caption{\small{Bayesian network related to the
joint pdf
$f(\V{y}, \V{a}, \V{b}, \V{m}, \V{z})$ in \eqref{eq-S:bayesian_factorization-3}.
For simplicity $S \! = \! 2$ \rd{sensors are} assumed, and the following short notations are used: 
$\protect\bar{\V{y}}_{k} \! \deq \rmv \V{y}_{n-1,k}$, 
$\protect\bar{\V{a}}^s \! \deq \rmv \V{a}^{(s)}_{n-1}$, 
$\protect\bar{\V{b}}^s \! \deq \rmv \V{b}^{(s)}_{n-1}$, 
$\protect\bar{M}^s \! \deq \rmv M^{(s)}_{n-1}$, 
$\protect\bar{\V{z}}_{m}^s \! \deq \rmv \V{z}^{(s)}_{n-1,m}$,
$\V{y}_{k} \! \deq \rmv \V{y}_{n,k}$, 
$\V{a}^s \! \deq \rmv \V{a}^{(s)}_{n}\rmv$, 
$\V{b}^s \! \deq \rmv \V{b}^{(s)}_{n}\rmv$, 
$M^s \! \deq \rmv M^{(s)}_{n}\rmv$, 
and $\V{z}^s_{m} \! \deq \! \V{z}^{(s)}_{n,m}$. 
The color of each edge matches the color of the associated parent node.}}
\label{fig:bayesianNetwork}
\vspace{3mm}
\end{figure}

As a first step, 
we express the joint pdf $f ( \V{y}, \V{a}, \V{b}, \V{m}, \V{z})$
in terms of the likelihood function and the prior pdf, i.e., 
\begin{align}
	\nn \\[-8mm]
	f(\V{y}, \V{a}, \V{b}, \V{m}, \V{z}) = f(\V{z} | \V{y}, \V{a}, \V{b}, \V{m} ) \ist f( \V{y},\V{a}, \V{b}, \V{m} ) = f(\V{z} | \V{y}, \V{a}, \V{b}, \V{m} ) \ist p( \V{a}, \V{b}, \V{m} | \V{y}) \ist f( \V{y}) \ist.
\label{eq-S:bayesian_factorization-0}
\end{align}
Using, in turn, Assumptions A1 and A2, we have 
\[
f( \V{y}) = f(\V{y}_{0}) \!\prod_{n' = 1}^{n} \!\rmv f(\V{y}_{n'} | \V{y}_{n' - 1} ) = \prod_{k = 1}^{K} \rmv f(\V{y}_{0,k}) \!\prod_{n' = 1}^{n} \!\rmv f(\V{y}_{n'\!,k} | \V{y}_{n' - 1,k} ) \ist,
\vspace{-1mm}
\]
and thus \eqref{eq-S:bayesian_factorization-0} 
\vspace{-1.5mm}
becomes
\[
f(\V{y}, \V{a}, \V{b}, \V{m}, \V{z}) = f(\V{z} | \V{y}, \V{a}, \V{b}, \V{m} ) \ist p( \V{a}, \V{b}, \V{m} | \V{y}) \prod_{k = 1}^{K} \rmv f(\V{y}_{0,k}) 
  \!\prod_{n' = 1}^{n} \!\rmv f(\V{y}_{n'\!,k} | \V{y}_{n' - 1,k} ) \ist.
\]
Next, Assumptions A3 and A4
give
\[
f(\V{y}, \V{a}, \V{b}, \V{m}, \V{z}) = f(\V{z} | \V{y}, \V{a}, \V{m} ) \Bigg( \prod_{k = 1}^{K} \rmv f(\V{y}_{0,k}) \Bigg) \!\prod_{n' = 1}^{n} 
  \!\Bigg( \prod_{s=1}^S  p\big(\V{a}_{n'}^{(s)}\rmv,\V{b}_{n'}^{(s)}\rmv, M_{n'}^{(s)} \big| \V{y}_{n'}\big) \rmv \Bigg) \prod_{k' = 1}^{K} \rmv f(\V{y}_{n'\!,k'} | \V{y}_{n' - 1,k'} ) \ist.
\]
Then, using
Assumptions A5, A6, and A7,
we obtain
\begin{align}
	f(\V{y}, \V{a}, \V{b}, \V{m}, \V{z}) &= \Bigg( \prod_{k = 1}^{K} \rmv f(\V{y}_{0,k}) \Bigg) \rmv\prod_{n' = 1}^{n} \!\Bigg( \prod_{s=1}^S \rmv f \big(\V{z}_{n'}^{(s)} \big| \V{y}_{n'}, \V{a}_{n'}^{(s)}\rmv, M_{n'}^{(s)} \big) \ist p\big(\V{a}_{n'}^{(s)}\rmv, \V{b}_{n'}^{(s)}\rmv, M_{n'}^{(s)} \big| \V{y}_{n'} \big) \rmv\Bigg) \nn \\[0mm]
	& \hspace{5mm} \times \prod_{k' = 1}^{K} \rmv f(\V{y}_{n'\!,k'} | \V{y}_{n' - 1,k'} ) \ist
	\label{eq-S:bayesian_network_fact} \\[0mm]
	&= \Bigg( \prod_{k = 1}^{K} \rmv f(\V{y}_{0,k}) \Bigg) \prod_{n' = 1}^{n}  \Bigg[ \prod_{s=1}^S \rmv \Bigg( \prod_{m=1}^{M_{n'}^{(s)}} \! f \big(\V{z}_{n'\rmv,m}^{(s)} \big| \V{y}_{n'}, \V{a}_{n'}^{(s)}\rmv, M_{n'}^{(s)} \big) \rmv \Bigg) \ist p\big(\V{a}_{n'}^{(s)}\rmv, \V{b}_{n'}^{(s)}\rmv, M_{n'}^{(s)} \big| \V{y}_{n'} \big) \Bigg] \nn \\[0mm]
	& \hspace{5mm} \times \prod_{k' = 1}^{K} \rmv f(\V{y}_{n'\!,k'} | \V{y}_{n' - 1,k'} ) \ist.
	\label{eq-S:bayesian_factorization-1} \\[-8mm]
\nn
\end{align}
We have 
\vspace{-2mm}
\begin{align}
p\big(\V{a}_{n}^{(s)}\rmv, \V{b}_{n}^{(s)}\rmv, M_{n}^{(s)} \big| \V{y}_{n} \big) 
&= p\big(\V{b}_{n}^{(s)} \big| \V{a}_{n}^{(s)}\rmv, M_{n}^{(s)}\rmv, \V{y}_{n}\big) \ist p\big(\V{a}_{n}^{(s)}\rmv, M_{n}^{(s)} \big| \V{y}_{n}\big) \nn\\[.5mm]
&= p\big(\V{b}_{n}^{(s)} \big| \V{a}_{n}^{(s)}\rmv, M_{n}^{(s)} \big) \ist p\big(\V{a}_{n}^{(s)}\rmv, M_{n}^{(s)} \big| \V{y}_{n}\big) \nn\\[.5mm]
&= p\big(\V{b}_{n}^{(s)} \big| \V{a}_{n}^{(s)}\rmv, M_{n}^{(s)} \big) \ist p\big(\V{a}_{n}^{(s)} \big| M_{n}^{(s)}\rmv,\V{y}_{n}\big) \ist p\big(M_{n}^{(s)} \big| \V{y}_{n} \big) \ist,
  \label{eq-S:bayesian_factorization-2} 
\end{align}
where Assumption A8 was used in the second step.
Inserting \eqref{eq-S:bayesian_factorization-2} into \eqref{eq-S:bayesian_factorization-1}, we finally obtain
\begin{align}
f(\V{y}, \V{a}, \V{b}, \V{m}, \V{z}) &=  \Bigg( \prod_{k = 1}^{K} \rmv f(\V{y}_{0,k}) \Bigg) \prod_{n' = 1}^{n}  \Bigg[ \prod_{s=1}^S 
  \rmv \Bigg( \prod_{m=1}^{M_{n'}^{(s)}} \! f \big(\V{z}_{n'\rmv,m}^{(s)} \big| \V{y}_{n'}, \V{a}_{n'}^{(s)}\rmv, M_{n'}^{(s)} \big) \rmv \Bigg) 
    \ist p\big(\V{b}_{n'}^{(s)} \big| \V{a}_{n'}^{(s)}\rmv, M_{n'}^{(s)} \big) \nn\\[0mm]
&\hspace{5mm} \times p\big(\V{a}_{n'}^{(s)} \big| M_{n'}^{(s)}\rmv,\V{y}_{n'}\big) \ist p\big(M_{n'}^{(s)} \big| \V{y}_{n'} \big) \Bigg] \prod_{k' = 1}^{K} \rmv f(\V{y}_{n'\!,k'} | \V{y}_{n' - 1,k'} ) \ist.
\label{eq-S:bayesian_factorization-3}
\end{align}
The Bayesian network illustrating the statistical dependencies of the random variables involved in \eqref{eq-S:bayesian_factorization-3}
is shown in Fig.~\ref{fig:bayesianNetwork}, assuming $S \! = \! 2$ \rd{sensors} for simplicity.

\section{Factorization \cite[Eq.~(3)]{GagSolBraMagMeyHla:J19} of the Joint Posterior pdf $f(\V{y},\V{a},\V{b}|\V{z})$}
\label{sec:Der_Joint_Post}

A detailed derivation of the factorization \cite[Eq.~(3)]{GagSolBraMagMeyHla:J19} of the joint posterior pdf $f(\V{y},\V{a}, \V{b} | \V{z})$ is provided in \cite{MeyBraWilHla:J17}.
For a sketch of that derivation, we first observe that the likelihood function $f \big(\V{z}_{n}^{(s)} \big| \V{y}_n, \V{a}_{n}^{(s)}\rmv, M_{n}^{(s)} \big)$ can be factorized as (cf. \cite{MeyBraWilHla:J17,SoldiMBH:J19})
\be
f \big(\V{z}_{n}^{(s)} \big| \V{y}_n, \V{a}_{n}^{(s)}\rmv, M_{n}^{(s)} \big)  =\ist C_w\big(\V{z}_{n}^{(s)}\rmv, M_{n}^{(s)}\big) 
\prod^{K}_{k=1} \rmv w\big( \V{x}_{n,k} , r_{n,k}, \ell_{n,k}, a^{(s)}_{n,k}; \V{z}_{n}^{(s)} \big) \ist, 
\label{eq-S:globalLikelihood_3}
\vspace{-1.5mm}
\ee
where $C_w\big(\V{z}_{n}^{(s)}\rmv, M_{n}^{(s)}\big) \triangleq \prod^{M_{n}^{(s)}}_{m = 1} f_{\text{FA}}\big( \V{z}^{(s)}_{n,m} \big)$ is a normalization factor 
that depends on $\V{z}_{n}^{(s)}$ and $M_{n}^{(s)}\rmv$, and $w \big(\V{x}_{n,k}, r_{n,k}, \ell_{n,k}, a^{(s)}_{n,k}; \V{z}_{n}^{(s)} \big)$ is defined 
\vspace{-3mm}
as
\begin{align}
w^{(s)} \big( \V{x}_{n,k}, 1, \ell_{n,k}, a^{(s)}_{n,k}; \V{z}_{n}^{(s)} \big) &\triangleq \begin{cases}
    \displaystyle\frac{f\big( \V{z}^{(s)}_{n,m} \big| \V{x}_{n,k}, \ell_{n,k} \big)}{f_{\text{FA}}\big( \V{z}^{(s)}_{n,m} \big)} \ist, 
      & \rmv\rmv a^{(s)}_{n,k} \!=\rmv m \rmv\in\rmv \cl{M}^{(s)}_n\\[2mm]
     1 \ist, & \rmv\rmv a^{(s)}_{n,k} \!=\rmv 0 ,
  \end{cases}\nn\\[.5mm]
w^{(s)} \big( \V{x}_{n,k}, 0, \ell_{n,k}, a^{(s)}_{n,k}; \V{z}_{n}^{(s)} \big) &\triangleq 1.
\label{eq-S:likelihoodFactors_w}
\end{align}
Furthermore, the prior probability mass function (pmf) $p\big(\V{a}_{n}^{(s)}\rmv, \V{b}_{n}^{(s)}\rmv, M_{n}^{(s)} \big| \V{y}_{n}\big)$ can be factorized as (cf. \cite{MeyBraWilHla:J17,SoldiMBH:J19})
\begin{align}
	p\big(\V{a}_{n}^{(s)}\rmv, \V{b}_{n}^{(s)}\rmv, M_{n}^{(s)} \big| \V{y}_{n}\big) 
	= C_h\big(M_{n}^{(s)}\big) \prod^{K}_{k=1} \rmv h\big( \V{x}_{n ,k} , r_{n ,k}, \ell_{n,k}, a^{(s)}_{n,k}; M^{(s)}_{n}\big) \prod_{m=1}^{M_{n}^{(s)}} \!\rmv \Psi\big(a_{n,k}^{(s)} \ist,b_{n,m}^{(s)}\big) \ist,
	\label{eq-S:factorization_prior3}
\end{align}
where
$C_h\big(M_{n}^{(s)}\big) \triangleq e^{-\mu^{(s)}} (\mu^{(s)})^{M_{n}^{(s)}} \! / M_{n}^{(s)}!$ is a normalization factor that depends on $M_{n}^{(s)}\rmv$, and 
$h\big( \V{x}_{n,k}, \linebreak r_{n,k}, \ell_{n,k}, a^{(s)}_{n,k}; M^{(s)}_{n} \big)$ is defined 
\vspace{-3.5mm}
as
\begin{align}
h^{(s)}\big( \V{x}_{n,k}, 1,  \ell_{n,k}, a^{(s)}_{n,k}; M^{(s)}_{n} \big) &\triangleq \begin{cases}
    \displaystyle \frac{P_{\text{d}}^{(s)}(\V{x}_{n,k},\ell_{n,k})}{\mu^{(s)}} \ist, 
      & \rmv\rmv a^{(s)}_{n,k} \rmv\in\rmv \cl{M}^{(s)}_n\\[3.5mm]
     1 \!-\rmv P_{\text{d}}^{(s)}(\V{x}_{n,k},\ell_{n,k}) \ist, & \rmv\rmv a^{(s)}_{n,k} = 0,
  \end{cases} \nn \\[1.5mm]
h^{(s)}\big( \V{x}_{n,k}, 0, \ell_{n,k}, a^{(s)}_{n,k}; M^{(s)}_{n} \big) &\triangleq \delta_{a^{(s)}_{n,k},0} \ist.
\label{eq-S:likelihoodFactors_h}
\end{align}
Then, we note that because the measurements $\V{z}$ are observed and thus fixed, $M^{(s)}_{n}$ and $\V{m}$ are fixed
as well, and thus we have $f ( \V{y},\V{a},\V{b} | \V{z} ) \! = \! f ( \V{y},\V{a},\V{b},\V{m} | \V{z} )$.
Using Bayes' rule, we further have up to a constant factor
\begin{align}
	f ( \V{y},\V{a},\V{b} | \V{z} ) \propto f ( \V{z} | \V{y},\V{a},\V{b},\V{m} ) f ( \V{y},\V{a},\V{b},\V{m} ) = f ( \V{y},\V{a},\V{b},\V{m},\V{z} )\ist.
\end{align}
Therefore, the factorization of the joint posterior pdf $f ( \V{y},\V{a},\V{b} | \V{z} )$ is proportional (up to a constant factor) to the factorization
of the joint pdf $f ( \V{y},\V{a},\V{b},\V{m},\V{z} )$.
Hence, by inserting
\eqref{eq-S:globalLikelihood_3} and 
\eqref{eq-S:factorization_prior3} into \eqref{eq-S:bayesian_network_fact}, and omitting the constants $C_w\big(\V{z}_{n}^{(s)}\rmv, M_{n}^{(s)}\big)$ and $C_h\big(M_{n}^{(s)}\big)$, 
we obtain the factorization 
\cite[Eq.~(3)]{GagSolBraMagMeyHla:J19}, 
\vspace{-1mm}
i.e., 
\[
	f(\V{y},\V{a},\V{b} |\V{z}) \propto \prod_{k=1}^K\rmv  f(\V{y}_{0,k}) \rmv \prod_{n'=1}^n \!\rmv f(\V{y}_{n'\!,k}|\V{y}_{n'-1,k} ) 
	   \prod_{s=1}^S  \rmv \upsilon^{(s)} \big(\V{x}_{n'\!,k},r_{n'\!,k}, \ell_{n'\!,k}, a_{n'\!,k}^{(s)}; \V{z}_{n'}^{(s)}\big)   
	   \rmv\prod_{m=1}^{M_{n'}^{(s)}} \! \Psi\big(a_{n'\!,k}^{(s)},b_{n'\!,m}^{(s)}\big) \ist,
\vspace{-1mm}
\]
where 
\vspace{-1mm}
\[
\upsilon^{(s)} \big(\V{x}_{n,k},r_{n,k}, \ell_{n,k}, a_{n,k}^{(s)}; \V{z}_{n}^{(s)}\big) \triangleq w^{(s)} \big( \V{x}_{n,k}, r_{n,k}, \ell_{n,k}, a^{(s)}_{n,k}; \V{z}_{n}^{(s)} \big) 
  h^{(s)}\big( \V{x}_{n,k}, r_{n,k},  \ell_{n,k}, a^{(s)}_{n,k}; M^{(s)}_{n} \big) \ist.
\vspace{1mm}
\]

\section{SPA-based Message Passing Method}
\label{sec_SM:SPAmethod}

\begin{figure}[t!]
\vspace{3mm}
\centering

\psfrag{f1}[c][c][.85]{\raisebox{0mm}{\hspace{0mm}$f_1$}}
\psfrag{fk}[c][c][.85]{\raisebox{0mm}{\hspace{0mm}$f_K$}}

\psfrag{ft1}[c][c][.65]{\raisebox{0mm}{\hspace{0mm}\rd{$\tilde{f}_1$}}}
\psfrag{ftk}[c][c][.65]{\raisebox{0mm}{\hspace{0mm}\rd{$\tilde{f}_K$}}}
\psfrag{ftm1}[c][c][.65]{\raisebox{0mm}{\hspace{-3mm}\rd{$\tilde{f}_1^-$}}}
\psfrag{ftmk}[c][c][.65]{\raisebox{0mm}{\hspace{-3mm}\rd{$\tilde{f}_K^-$}}}

\psfrag{y1}[c][c][.85]{\raisebox{-2mm}{\hspace{0mm}$\V{y}_1$}}
\psfrag{yk}[c][c][.85]{\raisebox{-2mm}{\hspace{0mm}$\V{y}_K$}}

\psfrag{v1}[c][c][.85]{\raisebox{0mm}{\hspace{0mm}$\upsilon_1$}}
\psfrag{vk}[c][c][.85]{\raisebox{0mm}{\hspace{0mm}$\upsilon_K$}}

\psfrag{a1}[c][c][.85]{\raisebox{0mm}{\hspace{0mm}$a_1$}}
\psfrag{ak}[c][c][.85]{\raisebox{0mm}{\hspace{0mm}$a_K$}}

\psfrag{b1}[c][c][.85]{\raisebox{0mm}{\hspace{0mm}$b_1$}}
\psfrag{bm}[c][c][.85]{\raisebox{0mm}{\hspace{0mm}$b_M$}}
\psfrag{p11}[c][c][.85]{\raisebox{0mm}{\hspace{0mm}$\Psi_{11}$}}
\psfrag{p1m}[c][c][.85]{\raisebox{4mm}{\hspace{-4mm}$\Psi_{1M}$}} 
\psfrag{pk1}[c][c][.85]{\raisebox{0mm}{\hspace{-4mm}$\Psi_{K1}$}}
\psfrag{pkm}[c][c][.85]{\raisebox{2mm}{\hspace{0mm}$\Psi_{KM}$}}

\psfrag{c11}[c][c][.65]{\raisebox{0mm}{\hspace{0mm}\rd{$\zeta_{1,1}$}}}
\psfrag{n11}[c][c][.65]{\raisebox{0mm}{\hspace{0mm}\rd{$\nu_{1,1}$}}}
\psfrag{nM1}[c][c][.65]{\raisebox{4mm}{\hspace{-2mm}\rd{$\nu_{M,1}$}}}
\psfrag{cK1}[c][c][.65]{\raisebox{-5mm}{\hspace{0mm}\rd{$\nu_{K,1}$}}}
\psfrag{n1K}[c][c][.65]{\raisebox{3mm}{\hspace{0mm}\rd{$\nu_{1,K}$}}}
\psfrag{c1M}[c][c][.65]{\raisebox{-3mm}{\hspace{0mm}\rd{$\zeta_{1,M}$}}}
\psfrag{nMK}[c][c][.65]{\raisebox{-3mm}{\hspace{-2mm}\rd{$\eta_{M,K}$}}}
\psfrag{cKM}[c][c][.65]{\raisebox{-3mm}{\hspace{3mm}\rd{$\zeta_{K,M}$}}}
\psfrag{be1}[c][c][.65]{\raisebox{0mm}{\hspace{0mm}\rd{$\beta_{1}$}}}
\psfrag{bek}[c][c][.65]{\raisebox{0mm}{\hspace{0mm}\rd{$\beta_{K}$}}}
\psfrag{et1}[c][c][.65]{\raisebox{-4mm}{\hspace{0mm}\rd{$\eta_{1}$}}}
\psfrag{etk}[c][c][.65]{\raisebox{-3mm}{\hspace{0mm}\rd{$\eta_{K}$}}}
\psfrag{al1}[c][c][.65]{\raisebox{-2mm}{\hspace{-1mm}\rd{$\alpha_{1}$}}}
\psfrag{alk}[c][c][.65]{\raisebox{-2mm}{\hspace{0mm}\rd{$\alpha_{K}$}}}
\psfrag{gm1}[c][c][.65]{\raisebox{-3mm}{\hspace{0mm}\rd{$\gamma_1$}}}
\psfrag{gmk}[c][c][.65]{\raisebox{-3mm}{\hspace{0mm}\rd{$\gamma_{K}$}}}
\psfrag{s1}[c][c][.85]{\raisebox{-4mm}{\hspace{-7mm}$s \! = \! 1$}}
\psfrag{s2}[c][c][.85]{\raisebox{-4mm}{\hspace{-7mm}$s \! = \! S$}}

\includegraphics[scale=.87]{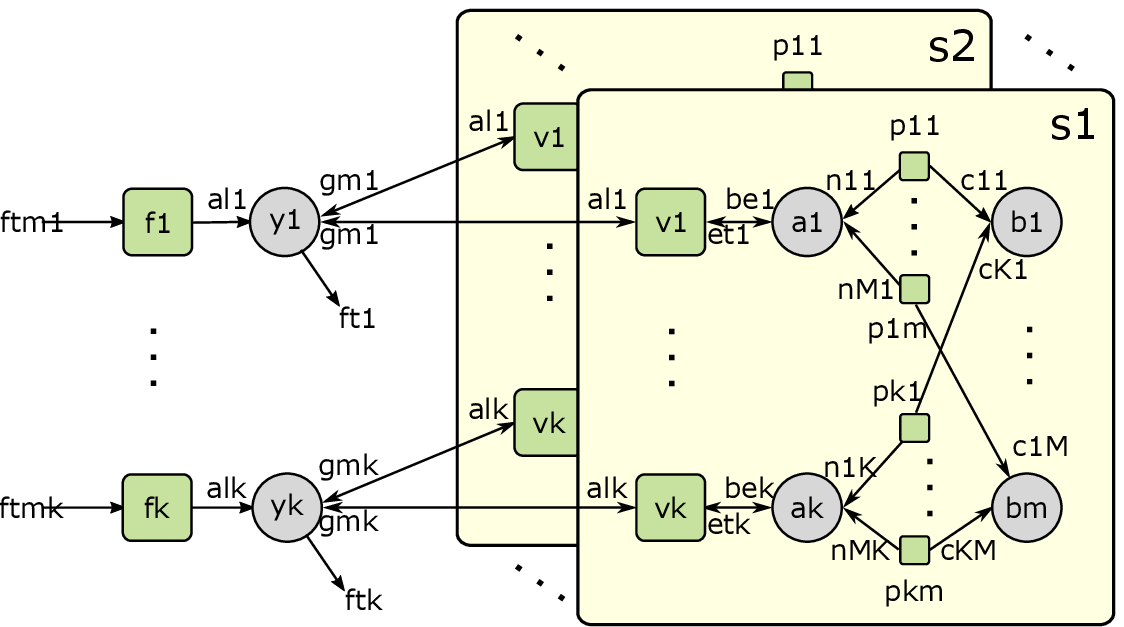}

\vspace{.7mm}

\caption{\small{Factor graph describing the factorization of $f(\V{y}, \V{a},\V{b} | \V{z})$ in \cite[Eq.~(3)]{GagSolBraMagMeyHla:J19} for one time step $n$. For simplicity, the sensor index $s$ and the time index $n$ are omitted, and the following short notations are used:
$f_{k} \! \triangleq \! f(\V{y}_{n,k}|\V{y}_{n - 1,k} )$,
$\upsilon_{k} \! \triangleq \rmv \upsilon^{(s)} \big(\V{y}_{n,k}, a_{n,k}^{(s)}; \V{z}_n^{(s)}\big)$,
$a_{k} \! \deq \rmv a_{n,k}^{(s)}$,
$b_{m} \! \deq \rmv b_{n,m}^{(s)}$,
$\Psi_{km} \! \triangleq \rmv \Psi_{km}\big(a_{n,k}^{(s)},b_{m,n}^{(s)}\big)$,
$M \! \triangleq \rmv M_n^{(s)}\rmv$,
$\tilde{f}^-_{k} \! \deq \rmv \tilde{f}(\V{y}_{n-1,k})$,
$\tilde{f}_{k} \! \deq \rmv \tilde{f}(\V{y}_{n,k})$,
$\alpha_{k} \! \triangleq \rmv  \alpha (\V{y}_{n,k})$,
$\beta_{k} \! \triangleq \!  \beta \big(a^{(s)}_{n,k}\big)$, 
$\eta_{k} \! \triangleq \rmv \eta \big(a^{(s)}_{n,k}\big)$, 
$\gamma_{k} \! \triangleq \rmv \gamma^{(s)} (\V{y}_{n,k})$, 
$\nu_{m,k} \! \triangleq \rmv   \nu^{(p)}_{m \rightarrow k}\big(a^{(s)}_{n,k}\big)$, and 
$\zeta_{k,m}\! \triangleq \rmv  \zeta^{(p)}_{k \rightarrow m}\big(b^{(s)}_{n,m}\big)$}.}
\label{fig-S:factorGraphMS}

\end{figure}

This section provides a description
of the classification-aided 
multitarget tracking method proposed in \cite[Sec.~III]{GagSolBraMagMeyHla:J19}.
Following the approach in \cite{MeyBraWilHla:J17,SoldiMBH:J19}, approximations of the marginal posterior pdfs 
$f(\V{y}_{n,k}|\V{z})\rrmv= \rrmv f(\V{x}_{n,k}, \linebreak r_{n,k}, \ell_{n,k} |\V{z})$, known as \emph{beliefs} and denoted as $\tilde{f}(\V{x}_{n,k},r_{n,k},\ell_{n,k})$, 
are
calculated at each time $n$ for all PTs $k$ in an efficient way by running the iterative sum-product algorithm (SPA) \cite{KscFreLoe:01,LoeDauHuKorPinKsc:J07}
on the factor graph depicted in Fig.~\ref{fig-S:factorGraphMS}. 
Since this factor graph contains loops, there is no unique order of calculating the individual messages passed between the nodes of the factor graph, and 
different orders may result in different beliefs. In our method, the order is defined by the following two rules: first, 
messages are not sent backward in time, and second, \emph{iterative} message passing is only performed for probabilistic data association, and 
separately at each time step and at each sensor. The second rule implies that for loops
involving different sensors, only a single 
message passing iteration is 
performed.

{\centering
  \begin{algorithm}[!t]
    \setstretch{.08}
 \scriptsize
\textbf{Input} (from previous time $n \!-\! 1$): 
$\tilde{f}(\V{x}_{n-1,k},r_{n-1,k}, \ell_{n-1,k})$
\vspace{2mm}

\textbf{Prediction}:
\For{
\rd{$k \! \in \! \{ 1,\ldots,K \}$}}{ 
\vspace{2mm}
\begin{align*}
&\alpha(\V{x}_{n,k},r_{n,k},\ell_{n,k} \! = \! i) \\
&\hspace{1mm}=\sum_{j=1}^C \int \! f(\V{x}_{n,k},r_{n,k}|\ell_{n,k} \! = \! i, \V{x}_{n-1,k},r_{n-1,k}\!=\!1) \, d_{i,j}(\V{x}_{n-1,k}) \ist\tilde{f}(\V{x}_{n-1,k},r_{n-1,k} \!=\!1, \ell_{n-1,k}\!=\!j) \,\mathrm{d}\V{x}_{n-1,k} \\
&\hspace{4mm}+\frac{1}{C} \int \! f(\V{x}_{n,k},r_{n,k}|\ell_{n,k} \! = \! i,\V{x}_{n-1,k},r_{n-1,k} \!=\!0) \,\tilde{F}(\V{x}_{n-1,k}) \,\mathrm{d}\V{x}_{n-1,k} \\
\end{align*}
\vspace{2mm}
for $i \! \in \! \{ 1,\ldots,C  \}$, and with \rd{$d_{i,j}(\V{x}_{n-1,k}) \! \deq \! [\V{D}(\V{x}_{n-1,k})]_{i,j}$ and} $\tilde{F}(\V{x}_{n-1,k}) \deq \sum_{j=1}^C \tilde{f}(\V{x}_{n-1,k},r_{n-1,k} \!=\! 0, \ell_{n-1,k}\!=\!j)$}

\vspace{2mm}

\For{\upshape all
\rd{$s \! \in \! \{ 1,\ldots,S \}$} (in parallel)}{ 
\vspace{2mm}

\textbf{Measurement evaluation}: \For{
\rd{$k \! \in \! \{ 1,\ldots,K \}$}}{
\vspace{2mm}

\begin{align*}
\beta\big( a_{n,k}^{(s)} \big) &= \!\! \sum_{\ell_{n,k} = 1}^C \int \! \upsilon^{(s)}\big(\V{x}_{n,k},
1,\ell_{n,k}, a_{n,k}^{(s)};\V{z}_n^{(s)}\big) \,\alpha(\V{x}_{n,k}, 1,\ell_{n,k}) \,\mathrm{d}\V{x}_{n,k} \ist + \delta_{a_{n,k}^{(s)},0} \ist \alpha_{n,k}\\[-.6mm]
\end{align*}
with $\alpha_{n,k} \triangleq \int \sum_{\ell_{n,k}=1}^C \alpha (\V{x}_{n,k}, 0,\ell_{n,k}) \,\mathrm{d}\V{x}_{n,k}$}
\vspace{2mm}

\textbf{Data association}:

\vspace{2mm}

\For{\upshape
\rd{$k \! \in \! \{ 1,\ldots,K \}$},
\rd{$m \! \in \! \{1,\ldots,M_{n}^{(s)} \}$}, and $p \!\in\! \{1,\dots,P\}$}{ 
\vspace{2mm}

\begin{align*}
\nu_{m \rightarrow k}^{(p)}(a_{n,k}^{(s)}) &= \hspace{-1mm} \sum_{b_{n,m}^{(s)} = 0}^K \hspace{-1.5mm} \Psi(a_{n,k}^{(s)},b_{n,m}^{(s)})
    \rd{\prod_{\substack{k' = 1 \\ k' \neq k}}^{K}}
    \zeta_{k' \rightarrow m}^{(p-1)}(b_{n,m}^{(s)}) \\
\zeta_{k \rightarrow m}^{(p)}(b_{n,m}^{(s)}) &= \hspace{-1mm} \sum_{a_{n,k}^{(s)} = 0}^{M_{n}^{(s)}} \hspace{-1mm} \beta(a_{n,k}^{(s)}) \ist \Psi(a_{n,k}^{(s)},b_{n,m}^{(s)})
    \rd{\prod_{\substack{m' = 1 \\ m' \neq m}}^{M_{n}^{(s)}}}
    \nu_{m'\rightarrow k}^{(p)}(a_{n,k}^{(s)}) \\[-1mm]
\end{align*}
with $\zeta_{k \rightarrow m}^{(0)}(b_{n,m}^{(s)}) = \sum_{a_{n,k}^{(s)} = 0}^{M_{n}^{(s)} } \beta(a_{n,k}^{(s)}) \ist \Psi(a_{n,k}^{(s)},b_{n,m}^{(s)})$}
\vspace{2mm}

\For{\upshape
\rd{$k \! \in \! \{ 1,\ldots,K \}$}}{ 
\vspace{2mm}

\begin{align*}
	\eta\big(a_{n,k}^{(s)}\big) = \prod_{m=1}^{M_{n}^{(s)}} \nu_{m \rightarrow k}^{(P)} \big(a_{n,k}^{(s)}\big)
\end{align*}
}
\vspace{2mm}

\textbf{Measurement update}: \For{
\rd{$k \! \in \! \{ 1,\ldots,K \}$}}{

\begin{equation*}
\gamma^{(s)}(\V{x}_{n,k},r_{n,k}, \ell_{n,k}) = \hspace{-1mm} \sum_{a_{n,k}^{(s)} = 0}^{M_{n}^{(s)}} \hspace{-1mm} \upsilon^{(s)}\big(\V{x}_{n,k},r_{n,k},\ell_{n,k},a_{n,k}^{(s)};\V{z}_n^{(s)}\big) \,\eta\big(a_{n,k}^{(s)}\big)
\vspace{-1mm}
\end{equation*}
}
\vspace{2mm}}
\vspace{2mm}

\textbf{Belief calculation}: \For{
\rd{$k \! \in \! \{ 1,\ldots,K \}$}}{
\vspace{2mm}
\begin{align*}
&\tilde{f}(\V{x}_{n,k},r_{n,k},\ell_{n,k}) = \frac{1}{C_{n,k}} \, \alpha(\V{x}_{n,k},r_{n,k},\ell_{n,k}) \rd{\prod_{s = 1}^{S}} \gamma^{(s)}(\V{x}_{n,k},r_{n,k}, \ell_{n,k}) \nn \\
\\[-3mm]
\end{align*}

\hspace{0mm} where $C_{n,k}$ is a normalization constant
}
\caption{\small{The proposed SPA-based message passing method---listing of the operations performed\vspace{.5mm} at time $n$.
Expressions of the pdfs $f(\V{x}_{n,k},r_{n,k}|\ell_{n,k},\V{x}_{n-1,k},r_{n-1,k} \!=\!0)$ \vspace{.7mm} and $f(\V{x}_{n,k},r_{n,k}|\ell_{n,k},\V{x}_{n-1,k},r_{n-1,k} \!=\!1)$ involved in the prediction step
are provided in \cite[Eq. (8)]{SoldiMBH:J19} and \cite[Eq. (9)]{SoldiMBH:J19}, respectively.}}
\label{alg:BPalgo}
\end{algorithm}
\par
}
Combining this message passing order with the generic SPA rules for calculating messages and beliefs \cite{KscFreLoe:01,LoeDauHuKorPinKsc:J07}, one obtains the 
SPA message passing operations constituting the proposed method. The calculation steps
performed at time $n$ are listed in Algorithm \ref{alg:BPalgo}. 
First, a \textit{prediction} step is performed for all PTs
\rd{$k \! \in \! \{1,\ldots,K \}$}, which comprises the calculation of messages 
$\alpha(\V{y}_{n,k}) \rmv=\rmv \alpha(\V{x}_{n,k}, r_{n,k}, \ell_{n,k})$.
\rd{This calculation involves the beliefs computed at the previous time $n-1$, i.e., $\tilde{f}(\V{x}_{n-1,k},r_{n-1,k}, \ell_{n-1,k})$, and the transition pdfs 
$f(\V{x}_{n,k}, r_{n,k}, \ell_{n,k} | \V{x}_{n-1,k}, r_{n-1,k}, \ell_{n-1,k})$.}
Next, the following steps are performed for all PTs
\rd{$k \! \in \! \{1,\ldots,K \}$} and for all sensors
\rd{$s \! \in \! \{1,\ldots,S \}$} in parallel:  
a \textit{measurement evaluation} step, in which \rd{the measurements, incorporated into the function $\upsilon^{(s)}\big(\V{x}_{n,k}, r_{n,k},\ell_{n,k},a_{n,k}^{(s)}; \V{z}_{n}^{(s)}\big)$, 
are employed to calculate} messages $\beta \big( a_{n,k}^{(s)} \big)$;
a \textit{data association} step, 
in which the messages $\beta \big( a_{n,k}^{(s)} \big)$ are converted into messages $\eta \big( a_{n,k}^{(s)} \big)$ \rd{by means an iterative procedure 
(in Algorithm \ref{alg:BPalgo}, $p \!\in\! \{1, \ldots, P \}$ denotes the iteration index)}; 
and a \textit{measurement update} step, in which messages $\gamma^{(s)}(\V{x}_{n,k},r_{n,k},\ell_{n,k})$ are calculated.
We note that the data association step closely follows \cite{WilLau:J14} and is equal to that in \cite{MeyBraWilHla:J17}.
Finally, in the \textit{belief calculation} step, beliefs  $\tilde{f}(\V{x}_{n,k},r_{n,k},\ell_{n,k})$ are calculated for all PTs
\rd{$k \! \in \! \{ 1,\ldots,K \}$} and used as input at 
the next time $n+1$. 
The detection of the individual PTs $k$ and the estimation of the states $\V{x}_{n,k}$ of the detected PTs
are carried out by using the beliefs $\tilde{f}(\V{x}_{n,k},r_{n,k}\!=\!1,\ell_{n,k})$ instead of the posterior pdfs $f(\V{x}_{n,k}, r_{n,k} \!=\! 1, \ell_{n,k} | \V{z})$ (cf. the discussion at the beginning of \cite[Sec.~III]{GagSolBraMagMeyHla:J19}). 
A particle-based implementation of this SPA-based message passing method that avoids an explicit evaluation of integrals and 
message products can be obtained by extending the 
implementation presented in \cite{MeyBraWilHla:J17}.

\section{Simulation Results}
\label{sec_SM:sim_results}

In this section, we analyze the performance of the proposed classification-aided 
multitarget tracking method for $C \rmv=\rmv 1,2,3,6$ target classes.
We recall that $C \rmv = \rmv 1$ means that all the targets fall into the same class,
and $C \rmv = 2,3,6$ means that the targets belong to $2$, $3$, or $6$ different classes.
In all cases, the classifier also distinguishes between 
target- and clutter-generated measurements.
The main purpose of the presented simulation results is to demonstrate the dependence of the 
performance of our method on the performance of the classifier, which is modeled by the confusion matrix $\V{G}^{(s)}(\V{x}_{n,k}) \! = \! \V{G}$ and the pmf 
$p_0\big(\zeta_{n,m}^{(s)} \big| \V{q}_{n,m}^{(s)}\big)$. 

The simulated scenario is that of \cite[Sec.~IV-A]{GagSolBraMagMeyHla:J19}, 
with the trajectories and the region of interest (ROI) shown in \cite[Fig.~2]{GagSolBraMagMeyHla:J19} and the mean number of false alarms set to $\mu^{(s)} \! = \! 20$.
For $C \!=\! 1$, all six targets belong to the same class $c\!=\! 1$; 
for $C \!=\rmv 2$, targets A, C, and E 
belong to class $c\!=\!1$ and targets B, D, and F belong to class $c\!=\!2$;
for $C \!=\rmv 3$, targets A and D belong to class $c \! = \! 1$, targets B and E belong to class  $c \! = \! 2$, and targets C and F belong to class $c \! = \! 3$; 
and for $C \!=\rmv 6$, targets A, B, C, D, E, and F belong to classes $c \! = \! 1$, $2$, $3$, $4$, $5$, and $6$, respectively.
We emphasize that the tracking method
has no prior knowledge about these
class affiliations. However, it uses the output of the classifier, which indicates---associated with each target-generated measurement produced by the sensor---an 
imperfect estimate of the target class, where the error of this estimate is modeled probabilistically by the confusion matrix $\V{G}$ (specified further below). 
The elements of the class transition matrix
\rd{$\V{D}(\V{x}_{n-1,k})\! = \! \V{D}$} are chosen as 
\rd{$[\V{D}]_{i,j} \! =\rmv 0.95$} if $i\!=\!j$ and
\rd{$[\V{D}]_{i,j} \! = \rmv 0.05/(C\!-\!1)$}
if $i \! \neq \! j$.
The performance
is assessed
in terms of
the time-averaged  MGOSPA error \cite{RahGarSven:C17}, the time-averaged MOSPA error \cite{SchVoVo:J08},
the time-averaged MOSPA-T error \cite{RisVoClarkVo:J11} (all three with order 1 and cutoff parameter 20 m; the time-averaged MOSPA-T error additionally 
with label error penalty 20 m),
and the false alarm rate (FAR), all averaged over 200 simulation runs.

We consider two cases. In the first case, for an increasing number of classes $C$, the diagonal elements of the confusion matrix $\V{G}$
are fixed whereas the off-diagonal elements are decreased. More specifically, we choose
\rd{$[\V{G}]_{i,j}  \!= \rmv 0.85$} if $i \! = \! j$ and
\rd{$[\V{G}]_{i,j} \!= \rmv 0.15/C$} if $i \! \neq \! j$, for $i \! \in \! \{ 0,1,\ldots,C \}$ and $j \! \in \! \{ 1,\ldots,C \}$. 
Furthermore, we choose $p_0\big(\zeta_{n,m}^{(s)} \big| \V{q}_{n,m}^{(s)}\big) \!=\rmv p_0\big(\zeta_{n,m}^{(s)}\big)$ 
as $p_0\big(\zeta_{n,m}^{(s)}\!=\!0\big)\!= \rmv 0.85$ and $p_0\big(\zeta_{n,m}^{(s)}\!=\!i\big) \!= \rmv 0.15/C$ for $i \! \in \! \{1,\ldots,C\}$.
Table~\ref{tab:fixed_diagonal} shows the MGOSPA, MOSPA, MOSPA-T, and FAR performance of the proposed method as well as of the baseline method of \cite{MeyBraWilHla:J17} (which does not use the classifier output) for $S \!=\! 1$ and $S \!=\rmv 2$ sensors. 
One can observe that, as expected, the performance of the proposed method improves for an increasing number of classes $C$, i.e., all the metrics---the MGOSPA, MOSPA, and MOSPA-T errors 
as well as the FAR---decrease as $C$ increases.

In the second case,
we keep the off-diagonal elements of the confusion matrix $\V{G}$ fixed whereas the
diagonal elements are decreased for increasing $C$.
More specifically, we choose
\rd{$[\V{G}]_{i,j} \!= \! 1-0.10 \rmv\cdot\rmv C$} if $i \! = \! j$ and
\rd{$[\V{G}]_{i,j}  \!= \! 0.10$} if $i \! \neq \! j$, for $i \! \in \! \{ 0,1, \ldots,C \}$ and $j \! \in \! \{ 1,\ldots,C \}$, as well as $p_0\big(\zeta_{n,m}^{(s)}\!=\!0\big)\!= \rmv 1-0.10 \rmv\cdot\rmv C$ and $p_0\big(\zeta_{n,m}^{(s)}\!=\!i\big) \!= \rmv 0.10$ for $i \! \in \! \{1,\ldots,C\}$.
Table~\ref{tab:fixed_off-diagonal} shows that for growing $C$, the
MOSPA-T error
of our method decreases
up to $C \!=\rmv 3$.
Indeed, here the diagonal elements of $\V{G}$ are still dominant, i.e., the classifier still provides sufficiently reliable estimates of the target classes.
However, for $C \!=\rmv 6$, the MOSPA-T error 
is larger, getting closer 
to that of the baseline method of \cite{MeyBraWilHla:J17}.
Here, the diagonal elements of $\V{G}$ are only
\rd{$[\V{G}]_{i,i} \!= \rmv 0.4$}, and the classifier is no longer able to reliably estimate the target classes.
Looking at the FAR, on the other hand,
one can observe that it consistently increases for growing $C$.
Indeed, as $C$ grows,
the ability of the classifier to correctly
identify
clutter-generated measurements decreases,
which results in a larger number of
false tracks 
among the estimated
tracks.
This also leads to higher MGOSPA and MOSPA errors.

\begin{table}[!t]
\vspace{0mm}
\renewcommand{\arraystretch}{1.2}
\scriptsize
\begin{subtable}{.48\linewidth}
\centering
\begin{tabular}{ c | c | c | c | c }
\hline
\multirow{2}{*}{$C$}	&	MGOSPA 	&	MOSPA	&	MOSPA-T &	FAR	\\
					&	[m]		&	[m]		&	[m]		&	[$\text{km}^{-2} \text{s}^{-1}$]		\\ \hline \hline
$1$					&	$27.8$	&	$6.5$	&	$11.2$	&	$1.83$	\\[-.5mm]
$2$					&	$25.6$	&	$5.9$	&	$7.7$	&	$1.24$	\\[-.5mm]
$3$					&	$24.2$	&	$5.6$	&	$6.0$	&	$1.01$	\\[-.5mm]
$6$					&	$22.9$	&	$5.2$	&	$5.3$	&	$0.80$	\\ \hline
Baseline\rmv\cite{MeyBraWilHla:J17}	&	$39.1$	&	$8.6$	&	$12.7$	&	$4.62$	\\ \hline
\end{tabular}
\caption{}
\label{tab:fixed_diagonal_s1}		
\end{subtable}
\hfill
\begin{subtable}{.48\linewidth}
\begin{tabular}{ c | c | c | c | c }
\hline
\multirow{2}{*}{$C$}	&	MGOSPA	&	MOSPA 	&  MOSPA-T	&	FAR	\\
					&   [m]		&	[m]		&	[m]		&	[$\text{km}^{-2} \text{s}^{-1}$]		\\ \hline \hline
$1$					&	$21.4$	&	$5.1$	&	$10.2$	&	$1.50$	\\[-.5mm]
$2$					&	$19.4$	&	$4.4$	&	$5.4$	&	$1.01$	\\[-.5mm]
$3$					&	$18.4$	&	$4.2$	&	$4.5$	&	$0.83$	\\[-.5mm]
$6$					&	$17.3$	&	$3.8$	&	$3.9$	&	$0.62$	\\ \hline
Baseline\rmv\cite{MeyBraWilHla:J17}	&	$30.0$	&	$7.1$	&	$11.6$	&	$3.69$	\\ \hline
\end{tabular}
\caption{}
\label{tab:fixed_diagonal_s2}
\end{subtable}
\vspace{-1.5mm}
\caption{\small{Time-averaged MGOSPA, MOSPA, and MOSPA-T errors as well as FAR for fixed diagonal elements of the confusion matrix $\V{G}$. 
(a) $S \!=\! 1$ \rd{sensor}, (b) $S \!=\! 2$ \rd{sensors}.}}	
\label{tab:fixed_diagonal}
\vspace{4mm}
\end{table}

\begin{table}[t!]

\vspace{0mm}

\renewcommand{\arraystretch}{1.3}
\scriptsize

\begin{subtable}{.48\linewidth}
\centering

\begin{tabular}{ c | c | c | c | c }
\hline
\multirow{2}{*}{$C$}	&	MGOSPA&	MOSPA	&	MOSPA-T 	&	FAR	\\
					&	[m]		&	[m]		&	[m]		&	[$\text{km}^{-2} \text{s}^{-1}$]		\\ \hline \hline
$1$					&	$25.0$	&	$5.7$	&	$10.6$	&	$1.15$	\\[-.5mm]
$2$					&	$27.5$	&	$6.4$	&	$8.6$	&	$1.69$	\\[-.5mm]
$3$					&	$29.4$	&	$6.8$	&	$8.1$	&	$2.19$	\\[-.5mm]
$6$					&	$35.9$	&	$8.1$	&	$11.1$	&	$3.78$	\\ \hline
Baseline\rmv\cite{MeyBraWilHla:J17}	&	$39.1$	&	$8.6$	&	$12.7$	&	$4.62$	\\ \hline
\end{tabular}
\caption{}
\label{tab:fixed_off-diagonal_s1}	
\end{subtable}\hfill
\begin{subtable}{.48\linewidth}

\begin{tabular}{ c | c | c | c | c }
\hline
\multirow{2}{*}{$C$}	&	MGOSPA &	MOSPA	&	MOSPA-T 	&	FAR	\\
					&	[m]		&	[m]		&	[m]		&	[$\text{km}^{-2} \text{s}^{-1}$]		\\ \hline \hline
$1$					&	$19.2$	&	$4.4$	&	$9.7$	&	$0.97$	\\[-.5mm]
$2$					&	$20.9$	&	$4.9$	&	$6.5$	&	$1.37$	\\[-.5mm]
$3$					&	$22.5$	&	$5.4$	&	$6.0$	&	$1.81$	\\[-.5mm]
$6$					&	$27.7$	&	$6.6$	&	$8.9$	&	$3.02$	\\ \hline
Baseline\rmv\cite{MeyBraWilHla:J17}	&	$30.0$	&	$7.1$	&	$11.6$	&	$3.69$	\\ \hline
\end{tabular}
\caption{}
\label{tab:fixed_off-diagonal_s2}
\end{subtable}
\vspace{-1.5mm}
\caption{\small{Time-averaged MGOSPA, MOSPA, and MOSPA-T errors as well as FAR for fixed off-diagonal elements of the confusion matrix $\V{G}$. 
(a) $S \!=\! 1$ \rd{sensor}, (b) $S \!=\! 2$ \rd{sensors}.}}	
\label{tab:fixed_off-diagonal}
\vspace{2mm}
\end{table}

\selectfont
\bibliographystyle{IEEEtran}
\bibliography{../../Bibliography/IEEEabrv,../../Bibliography/myBib,../../Bibliography/Wgroup,../../Bibliography/BiblioCV,../../Bibliography/Temp,../../Bibliography/biblio}